\documentclass{cpbtex}
\usepackage{graphicx}
\usepackage{booktabs} 
\usepackage{amsmath}

\begin{document}

\title{Rydberg-Mediated Nonlinear Quantum Optics}	

\author{Yun-Hui He(何云辉)$^{1,2}$,\ Chang-Cheng Li(李昌成)$^{1}$, \ Xu Shen(沈旭)$^{1}$,\\
\ Jing-Xu Bai(白景旭)$^{1,2}$,\ Xiao-Feng Shi(施小锋)$^{3,}$\thanks{Corresponding author. E-mail:xshi@hainanu.edu.cn}, \ Lin Li(李霖)$^{4,}$\thanks{Corresponding author. E-mail:li$\_$lin@hust.edu.cn}, \\ 
\ Yue-Chun Jiao(焦月春)$^{1,2,}$\thanks{Corresponding author. E-mail:ycjiao@sxu.edu.cn}, and Jian-Ming Zhao(赵建明)$^{1,2,}$\\
$^{1}${State Key Laboratory of Quantum Optics Technologies and Devices,}\\ {Institute of Laser Spectroscopy, Shanxi University, Taiyuan 030006, China}\\ 
$^{2}${Collaborative Innovation Center of Extreme Optics, Shanxi University, Taiyuan 030006, China}\\
$^{3}${Center for Theoretical Physics and School of Physics and Optoelectronic Engineering,}\\{Hainan University, Haikou 570228, China}\\	
$^{4}${MOE Key Laboratory of Fundamental Physical Quantities Measurement, Hubei Key Laboratory}\\{of Gravitation and Quantum Physics, PGMF, Institute for Quantum Science and Engineering,}\\{School of Physics, Huazhong University of Science and Technology, Wuhan 430074, China}}

\date{}
\maketitle

\begin{abstract}
Rydberg atoms have emerged as a versatile platform for quantum optics due to their exaggerated properties, particularly their strong long-range interactions, which enable a new regime of light–matter interaction. By mapping the interactions between Rydberg atoms onto photons, effective photon–photon interactions can be realized at the single-photon level, thereby overcoming the intrinsic weakness of conventional optical nonlinearities. In this review, we first introduce the fundamental physical principles of Rydberg-mediated quantum optics, and then discuss some key developments, including single-photon engineering, photonic quantum gates, contactless nonlinear optics, and quantum entanglement, providing a comprehensive overview of the current state and prospects of this rapidly developing field.

\end{abstract}
		
		

\section{Introduction}
\begin{figure}[htbp]
\centering  
\includegraphics[width=1\linewidth]{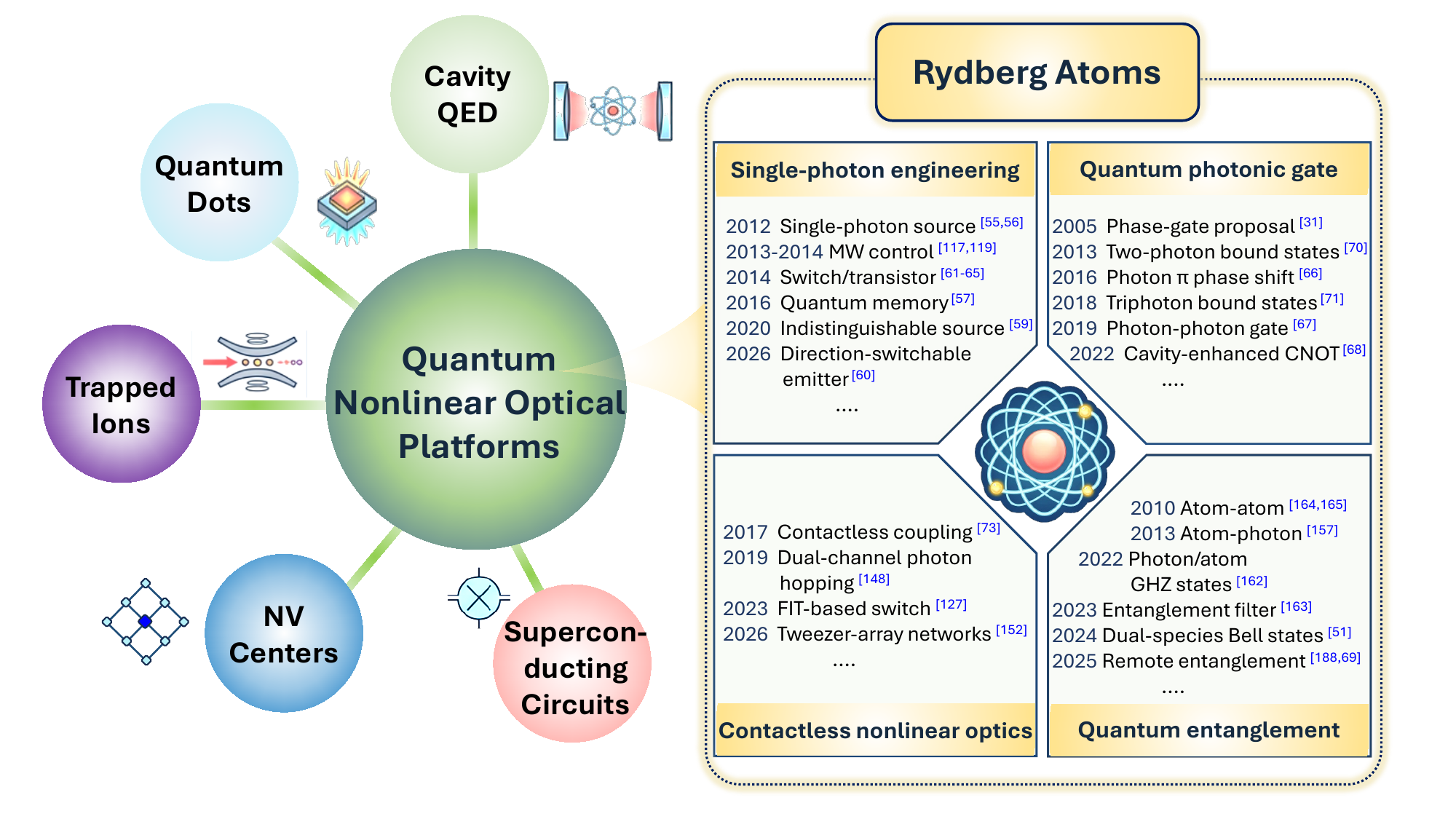}
\caption{Position of Rydberg-mediated nonlinear quantum optics within the broader landscape of quantum nonlinear optical platforms. Representative platforms, including cavity QED, quantum dots, trapped ions, NV centers, and superconducting circuits, are illustrated on the left. The right panel summarizes representative milestones in the development of Rydberg-mediated nonlinear quantum optics over the past two decades, highlighting advances in single-photon engineering, quantum photonic gates, contactless nonlinear optics, and quantum entanglement.}
\label{roadmap}
\end{figure}

Quantum nonlinear optics aims to realize strong interactions between individual photons, providing a foundation for quantum information processing, quantum networking, and many-body photonic physics. Over the past two decades, a variety of physical platforms have been developed to engineer optical nonlinearities at the quantum level, including cavity quantum electrodynamics (QED)\ucite{mabuchi2002cavity,birnbaum2005photon,dayan2008photon,kimble2008quantum,chen2013alloptical,reiserer2015cavitybased}, quantum dots\ucite{zwanenburg2013silicon,lu2021quantumdot, lodahl2015interfacing,senellart2017highperformance}, trapped ions\ucite{blatt2008entangled,monroe2013scaling,bruzewicz2019trappedion}, nitrogen-vacancy (NV) centers\ucite{doherty2013nitrogenvacancy,zhou2014quantum,awschalom2018quantum,song2019generation, ruf2021quantum}, superconducting circuits\ucite{devoret2013superconducting,blais2021circuit,gu2017microwave}, and Rydberg atoms\ucite{saffman2010quantum,firstenberg2016nonlinear,adams2020rydberg} (Fig.~\ref{roadmap}). Although these platforms have enabled remarkable advances in quantum optics, they generally involve trade-offs among interaction strength, coherence, scalability, operating conditions, and compatibility with propagating optical photons. Among them, Rydberg atoms, characterized by their highly excited electronic states, exhibit extraordinarily strong and long-range interactions that exceed those of ground-state atoms by several orders of magnitude\ucite{gallagher1994rydberg,saffman2010quantum,firstenberg2016nonlinear,sibalic2018rydberg,adams2020rydberg,shao2024rydberg}. A key breakthrough in Rydberg-mediated nonlinear quantum optics\ucite{firstenberg2016nonlinear} arises from the ability to transfer these strong atomic interactions to propagating light fields. By employing a Rydberg electromagnetically induced transparency (EIT) scheme, in which probe and coupling fields coherently connect the ground state to a highly excited Rydberg level, the strong Rydberg–Rydberg interaction can be mapped onto photons via the formation of Rydberg dark-state polaritons\ucite{fleischhauer2000darkstate,fleischhauer2002quantum,fleischhauer2005electromagnetically}, enabling effective photon–photon interactions at the single-photon level\ucite{friedler2005longrange,gorshkov2011photonphoton}. Such strong optical nonlinearities at a few photons are fundamentally inaccessible in conventional nonlinear optical media, where nonlinear responses typically require intense laser fields\ucite{chang2014quantum,boyd2020nonlinear}. 

A defining consequence of these interactions is the Rydberg blockade effect\ucite{lukin2001dipole,tong2004local,gaetan2009observation,pritchard2010cooperative,kazemi2023drivendissipative}, which prevents multiple Rydberg excitations within a characteristic volume. Under this condition, an entire atomic ensemble behaves as a single functional unit known as a ``superatom"\ucite{vuletic2006when,heidemann2007evidence,paris-mandoki2017freespace,kumlin2023quantum}. Within a superatom, the optical excitation is no longer shared by a single localized atom but is instead distributed among all the $N$ atoms as a coherent collective state\ucite{heidemann2007evidence,dudin2012observation}. This collective nature leads to a significant enhancement of the light-matter coupling strength, which scales as $\sqrt{N}$ as compared to the single-atom case, enabling high-efficiency mapping between photons and atomic excitations\ucite{wilk2010entanglement,saffman2016quantum}. 

By leveraging this superatom framework and EIT, Rydberg-mediated nonlinear optics has evolved from fundamental proof-of-principle studies to a versatile toolkit for quantum engineering.  Researchers have laid the foundation for engineering deterministic quantum states\ucite{zeng2017entangling, madjarov2020highfidelity,levine2018highfidelity,levine2019parallel,anand2024dualspecies} and high-fidelity multi-qubit gates\ucite{graham2022multiqubit, evered2023highfidelity, Bluvstein2024}, enabling the transition from fundamental atomic studies to the sophisticated manipulation of non-classical light. The ability to engineer such nonlinearities has opened a vast frontier of applications, including the realization of deterministic single-photon sources\ucite{pritchard2010cooperative,dudin2012Strongly,peyronel2012quantum,li2016Quantuma,petrosyan2018deterministic,ornelas-huerta2020Ondemand,li2026}, all-optical switches\ucite{baur2014SinglePhotonb}, and single-photon transistors\ucite{gorniaczyk2014SinglePhoton,tiarks2014SinglePhoton,gorniaczyk2016enhancement,liao2025nonlocal}. These systems have also enabled high-fidelity photon-photon quantum logic gates\ucite{tiarks2016optical,tiarks2019Photon,stolz2022quantumlogic}, which are essential for scalable optical quantum computing and long-distance quantum repeaters\ucite{kimble2008quantum,an2025entangling}. Beyond discrete logic, the effective interaction between photons allows for the synthesis of exotic states of light, such as photonic molecules and bound states\ucite{firstenberg2013attractive,liang2018observation}. Furthermore, Rydberg-mediated optics provides a robust platform for studying many-body physics, enabling the simulation of strongly correlated quantum phases and topological states of light\ucite{otterbach2013wigner,busche2017Contactlessb}. These developments signify a shift from fundamental atomic physics to a new paradigm of quantum engineering, where Rydberg systems serve as a versatile toolkit for manipulating non-classical light. Fig.~\ref{roadmap} highlights the historical development of key achievements in the Rydberg platform.

This review is organized to provide a systematic progression from fundamental principles to the latest experimental frontiers. In Section 2, we discuss the fundamentals of Rydberg physics, focusing on the scaling laws, the mechanism of Rydberg blockade, and the implementation of collective excitations via Rydberg-EIT and slow-light storage. In Section 3, we focus on Rydberg nonlinear quantum optics, specifically exploring single-photon engineering, the development of quantum photonic gates, contactless nonlinear optics, and the generation of quantum entanglement within these systems. Finally, in Section 4, we provide an outlook on the future of this rapidly evolving field.

\section{Fundamentals of Rydberg Physics Relevant to Quantum Optics}\label{fundamental}

\subsection{Properties of Rydberg atoms}

Rydberg atoms, defined by their high principal quantum number $n$, possess exaggerated physical properties that bridge the gap between classical and quantum mechanics. As summarized in the Tab.~\ref{Properties}, these fundamental properties follow dramatic scaling laws with respect to $n$. The electronic wavefunctions of these states extend over vast distances, with the orbit radius scaling as $r_n \propto n^2$. For a state with $n=60$, the atomic size can reach several hundred nanometers, thousands of times larger than a ground-state atom. This spatial expansion significantly reduces the overlap between the Rydberg state and lower-lying states, leading to a suppressed spontaneous emission rate and a corresponding increase in the radiative lifetime ($\tau \propto n^3$). Furthermore, the large size results in enormous electric dipole moments ($\mu \propto n^2$) and extreme polarizabilities ($\alpha \propto n^7$).
\begin{table}[htp!]
\centering
\small
\renewcommand{\arraystretch}{1} 
\setlength{\tabcolsep}{1mm} 
\begin{tabular}{ccc}
\toprule
Property & Symbol & $n$ Scaling \\
\midrule
Binding energy & $E_n$   & $n^{-2}$ \\
Level spacing & $\Delta E_n$   & $n^{-3}$ \\
Orbit radius & $r_n$   & $n^2$ \\
Lifetimes & $\tau$   & $n^3$ \\
Dipole moment & $\mu$   & $n^2$ \\
Polarizability & $\alpha$   & $n^7$ \\
Dipole-dipole interaction coefficient& $C_3$   & $n^4$ \\
van der Waals interaction coefficient& $C_6$   & $n^{11}$ \\
\bottomrule
\end{tabular}
\caption{Properties of Rydberg atoms.}
\label{Properties}
\end{table}
\begin{figure}[htbp]
\centering  
\includegraphics[width=0.6\linewidth]{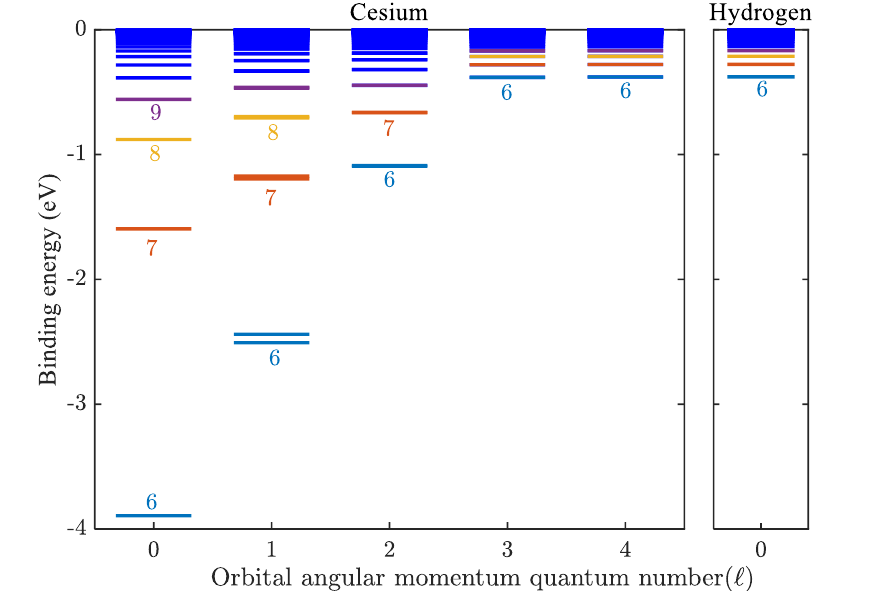}
\caption{Comparison of the energy level diagrams of the theoretically calculated Cesium atom (left panel) and hydrogen atom (right panel), where the principal quantum number is $n\in[6,60]$. These calculations are based on alkali Rydberg calculator (ARC) toolbox\ucite{robertson2021arc}.}
\label{property}
\end{figure}

A crucial practical distinction arises when considering alkali atoms, such as the Cesium atom illustrated in Fig.~\ref{property}. Due to the penetration and polarization of the ionic core by the valence electron, the energy levels of low-angular-momentum states ($\ell \leq 3$) are significantly shifted downward\ucite{low2012experimental}. This effect is quantified by the quantum defect $\delta_{\ell}$. In these realistic atomic systems, the aforementioned hydrogenic scaling laws remain remarkably robust if one simply replaces the principal quantum number $n$ with the effective principal quantum number $n^* = n - \delta_{\ell}$. As evidenced by the convergence of energy levels in Fig.~\ref{property}, for high-$\ell$ states where the electron remains far from the core ($\delta_{\ell} \to 0$), the manifold becomes nearly identical to that of a hydrogen atom and $n^*$ converges to $n$. In the following discussions, we use $n$ to denote the principal quantum number for simplicity, with the understanding that for alkali atoms, the effective value $n^*$ should be employed to account for quantum defects.

\begin{figure}[htbp]
\centering  
\includegraphics[width=0.65\linewidth]{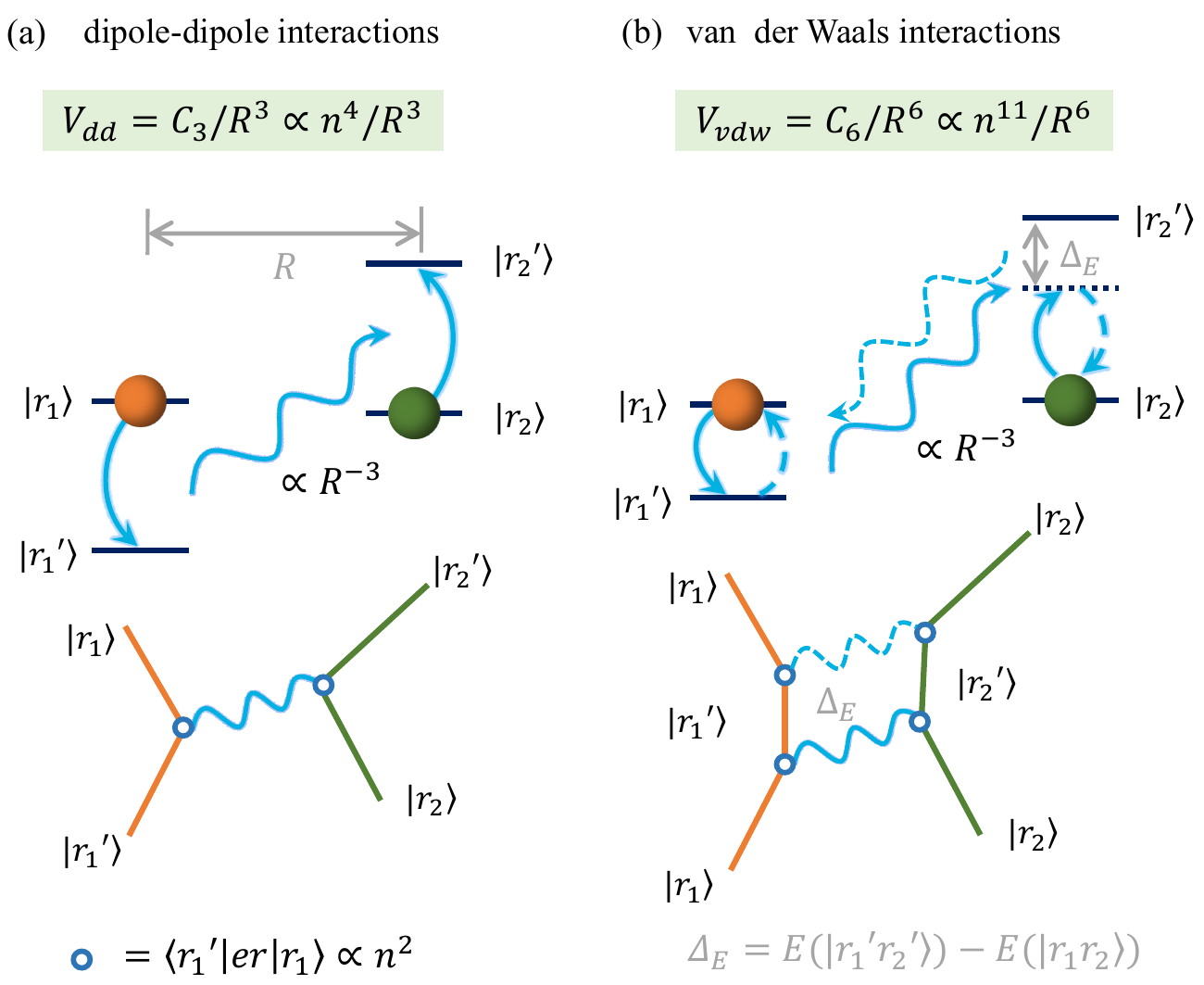}
\caption{Schematic diagram of the interactions between Rydberg states. (a) Dipole–dipole and (b) van der Waals interactions. Top panels: energy-level schemes for the two regimes. Bottom panels: Feynman diagrams representing the virtual photon exchange processes, highlighting the physical origins of the different scaling laws.}
\label{DD_VDW_model}
\end{figure}

The most defining feature of quantum optics is the strong Rydberg–Rydberg interaction. The exaggerated transition dipole moments of Rydberg atoms lead to exceptionally strong electrostatic interactions, which can be understood by considering two atoms separated by a distance $R$. The interaction is fundamentally governed by the dipole-dipole coupling, with a potential 
\begin{equation}\label{V(R)}
    V(R) = \frac{\mu_1\cdot \mu_2 }{R^3}-\frac{3(\mu_1\cdot R)(\mu_2\cdot R) }{R^5} .
\end{equation}
The specific physical manifestation of this coupling depends on the energy defect between the initial pair-state ($\lvert{r_1}\rangle,\lvert{r_2}\rangle$) and the accessible target states ($\lvert{r_1'}\rangle,\lvert{r_2'}\rangle$), i.e., $\Delta_E = E(\lvert{r_1'r_2'}\rangle) - E(\lvert{r_1r_2}\rangle)$. In the literature, the interaction between two Rydberg atoms is usually approximated as dipole-dipole (dd) interaction or van der Waals (vdW) interaction, while in practice there is no clean pure resonant dipole-dipole interaction because there is always non-resonant process\ucite{PhysRevA.77.032723}. On the other hand, only when $R\rightarrow\infty$ can the interaction be purely vdW\ucite{shi_quantum_2022}. Nonetheless, for the blockade mechanism to work, it is not necessary to have either interaction to be pure, but a hybrid of both is enough as long as no accidental zero interaction arises\ucite{PhysRevA.77.032723}.

When the atoms are in a resonant configuration, known as F\"{o}rster resonance ($\Delta_E \approx 0$), the interaction is dominated by the first-order exchange of a single virtual photon [see Fig.~\ref{DD_VDW_model}(a)]. In this regime, the potential follows a long-range resonant dipole-dipole form, $V_{dd} = C_3/R^3$. Given that the dipole moment scales as $\mu \propto n^2$, the interaction coefficient $C_3$ exhibits a scaling of $n^4$. This resonant coupling is characterized by a high degree of spatial anisotropy and provides the strongest available mechanism for fast quantum logic operations\ucite{barredo2015coherent,labuhn2016tunable,khazali2019polariton}.

Conversely, in most off-resonant experimental scenarios where $\Delta_E$ is large, the first-order interaction vanishes due to the energy mismatch. The interaction then proceeds via a second-order process involving the exchange of two virtual photons, as shown in Fig.~\ref{DD_VDW_model}(b). In this limit, the system is described by the vdW potential, $V_{vdW} = C_6/R^6$. By applying second-order perturbation theory, the coefficient can be expressed as $C_6 =(C_3)^2 / \Delta_E$. Taking into account the scaling of the dipole moments and the energy spacing ($\Delta_E \propto n^{-3}$), the vdW coefficient exhibits a remarkably steep scaling of $C_6 \propto n^{11}$. The transition between these two regimes occurs at a characteristic distance $R_{vdW} = |C_6/\Delta_E|^{1/6}$, where the interaction strength becomes comparable to the energy defect.

\subsection{Rydberg blockade and collective excitations}

Rydberg blockade\ucite{lukin2001dipole,tong2004local,gaetan2009observation,pritchard2010cooperative} is a phenomenon in which the strong interaction-induced energy shift produced by a single Rydberg excitation prevents the excitation of neighboring atoms within a characteristic blockade radius $R_b$. This mechanism forms the physical basis for achieving strong optical nonlinearities at the single-photon level and plays a central role in neutral-atom quantum information processing (QIP) and quantum simulation\ucite{urban2009observation,wilk2010entanglement,honer2010collective,pohl2010dynamical,weimer2010rydberg,wu2021concise}. In the blockade regime, the medium exhibits a highly nonlinear response, allowing only a single Rydberg excitation within a blockade volume, which serves as the fundamental mechanism underlying single-photon sources and Rydberg-mediated nonlinear optics\ucite{dudin2012Strongly,jiao2020singlephoton}.

\begin{figure}[htbp]
\centering  
\includegraphics[width=0.8\linewidth]{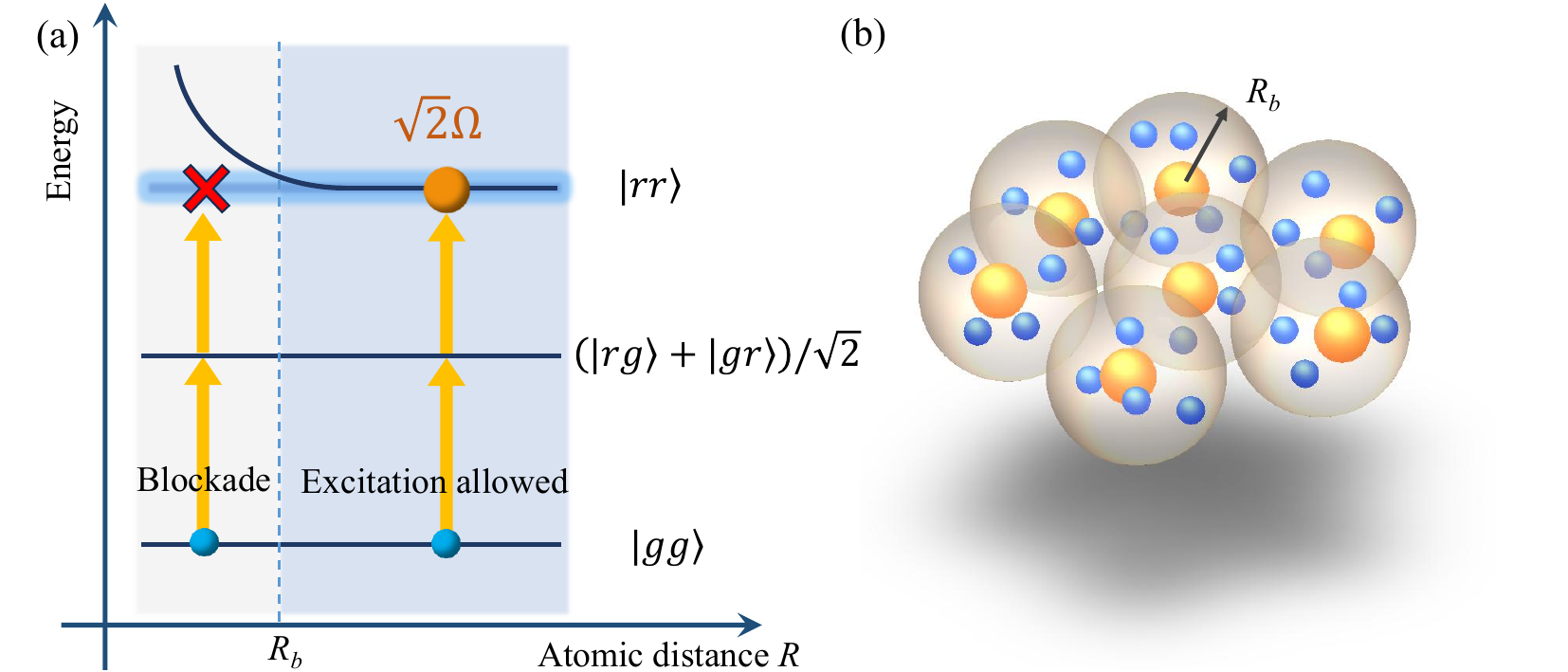}
\caption{(a) Schematic diagram of Rydberg blockade effect in the two-atom case. The blockade radius $R_b$ is defined as the characteristic distance at which the interaction-induced energy shift exceeds the excitation linewidth. Double excitation is therefore suppressed for $R<R_b$, whereas it is allowed for $R>R_b$. (b) Illustration of a Rydberg superatom. Atoms enclosed within a blockade sphere collectively share a single Rydberg excitation.} 
\label{blockade}
\end{figure}

To illustrate the underlying mechanism of the Rydberg blockade, we consider a simple two-atom system as depicted in Fig.~\ref{blockade}(a). When an atom is driven to a Rydberg state, its presence shifts the energy levels of nearby atoms through long-range interactions. When this interaction-induced shift exceeds the excitation linewidth dominated by the laser's Rabi frequency and power broadening, the laser becomes off-resonant for any further excitations. To characterize the spatial extent of this effect, the blockade radius $R_b$ is defined as the distance at which the interaction energy equals the excitation linewidth, typically characterized by the Rabi frequency $\Omega$, i.e., $|V(R_b)| = \hbar \sqrt{2}\Omega$. In the vdW regime $V(R)\propto C_6/R^6$, the blockade radius is given by $R_b = |C_6/(\hbar\sqrt{2} \Omega)|^{1/6}$\ucite{honer2011artificial,shao2024rydberg}. Given that the dispersion coefficient $C_6$ scales as $n^{11}$, the blockade radius is exquisitely sensitive to the principal quantum number $n$, allowing for precise experimental control. Physically, the blockade radius defines the characteristic interaction volume (blockade sphere), within which only one Rydberg excitation can be supported and all enclosed atoms collectively share the excitation.

In ensembles where the atomic cloud size is smaller than the blockade radius, the atoms no longer act as independent particles but instead behave as a single ``superatom", as illustrated in Fig.~\ref{blockade}(b). In this regime, the system is restricted to a two-level transition between the collective ground state $|G\rangle = |g_1 g_2 \dots g_N\rangle$ and a collective singly-excited Rydberg state $|R\rangle = \frac{1}{\sqrt{N}} \sum_{j=1}^N |g_1 \dots r_j \dots g_N\rangle$. This collective mapping leads to collective enhancement, where the transition between $|G\rangle$ and $|R\rangle$ occurs with an effective Rabi frequency $\Omega_{\text{eff}} = \sqrt{N} \Omega$. This collective enhancement enables efficient and coherent mapping between photonic excitations and collective Rydberg states, forming the basis for high-efficiency photon storage and strong photon–photon interactions in Rydberg-EIT systems.

\subsection{Rydberg EIT, dark-state polariton and photon storage}

\begin{figure}[htbp]
\centering  
\includegraphics[width=0.75\linewidth]{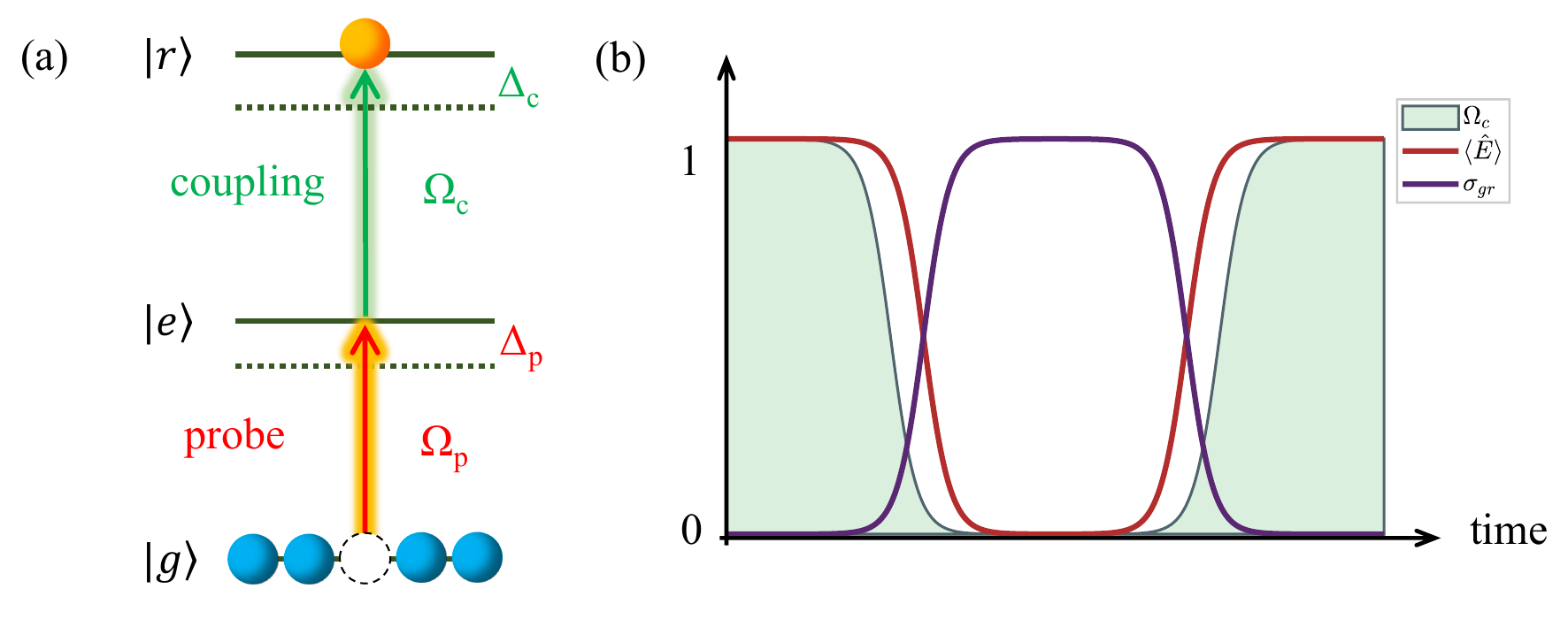}
\caption{(a) Rydberg-EIT level scheme. A weak probe field ($\Omega_p$) and a strong coupling field ($\Omega_c$) dress the ground state $|g\rangle$, intermediate state $|e\rangle$, and  Rydberg state $|r\rangle$. $\Delta_{p(c)}$ are the detunings of probe (coupling) laser. (b) Photon storage time sequence. As the control field $\Omega_c(t)$ (shaded green) is adiabatically turned off, the stored probe field $\langle\hat{E}\rangle$ (red) is transferred and frozen as a collective atomic coherence $\sigma_{gr}$ (purple), forming a dark-state polariton. The reverse process occurs upon switching $\Omega_c$ back on for retrieval.}
\label{EIT&polarton}
\end{figure}

While the Rydberg blockade provides the essential interaction mechanism, its practical implementation in optical systems often relies on EIT, which serves as a coherent interface between photons and collective atomic excitations\ucite{fleischhauer2005electromagnetically,mohapatra2007coherent,weatherill2008electromagnetically,pritchard2010cooperative,petrosyan2011electromagnetically,dudin2012Strongly}. The basic configuration is a three-level ladder-type system, as shown in Fig.~\ref{EIT&polarton}(a). A weak probe field drives the transition $|g\rangle \rightarrow |e\rangle$ with Rabi frequency $\Omega_p$ and detuning $\Delta_p$, while a strong control field with Rabi frequency $\Omega_c$ and detuning $\Delta_c$ couples the intermediate state $|e\rangle$ to a high-lying Rydberg state $|r\rangle$. The destructive interference between two excitation pathways suppresses resonant absorption and renders the medium transparent to the probe field.

Beyond transparency, the EIT medium exhibits a steep dispersion profile\ucite{fleischhauer2005electromagnetically}, resulting in a dramatically reduced group velocity of the probe pulse, a phenomenon commonly referred to as the slow-light effect\ucite{budker1999nonlinear,novikova2012electromagnetically}. The group velocity can be approximated as 
\begin{equation}
v_g = \frac{d\omega_p}{dk_p} \approx \frac{c}{{n}_g},
\end{equation}
where $\omega_p$ and $k_p$ denote the probe angular frequency and wave vector, respectively. $c$ is the speed of light and ${n}_g$ is the group index. The group velocity is reduced as the group index
\begin{equation}\label{eq:ng}
{n}_g = \frac{6\pi \rho_{N} c}{k_p^2} \frac{\gamma_e}{\Omega_c^2 + \gamma_r\gamma_e/4}
\end{equation}
increases. Here, $\rho_N$ is the atomic density and $\gamma_e$ ($\gamma_r$) is the decay rate of the intermediate (Rydberg) state. According to the above equations, the group velocity $v_g$ is determined by two key experimental parameters: the atomic density $\rho_N$ and the control field Rabi frequency $\Omega_c$. 

The atomic density $\rho_N$ is parameterized by the optical depth (OD) with $\text{OD} = \sigma_0 \rho_N L$, where $\sigma_0$ is the resonant absorption cross-section and $L$ is the length of the atomic medium. In the Rydberg blockade regime, this is further refined as the optical depth per blockade radius, $\text{OD}_b = \sigma_0 \rho_N R_b$. Achieving $\text{OD}_b \gtrsim 1$ is a prerequisite for mapping the strong Rydberg-Rydberg interactions onto the optical field, enabling single-photon level nonlinearities. In optically thick media with large OD, probe group velocities can be reduced to speeds as low as $17$~m/s or even lower\ucite{kash1999ultraslow,hau1999light}. 

On the other hand, the dynamic tunability of $\Omega_c$ enables the storage of light. By adiabatically ramping down $\Omega_c(t)$ to zero, the photonic component of the dark-state polariton (DSP) is completely converted into a stationary collective atomic excitation, as shown in Fig.~\ref{EIT&polarton}(b). The mixing angle $\theta(t)$ defines the hybrid nature of the DSP:
\begin{equation}
   \cos \theta(t) = \frac{\Omega_c(t)}{\sqrt{\Omega_c^2(t)+g^2N}}, \quad\tan \theta(t) = \frac{g\sqrt{N}}{\Omega_c(t)}, 
\end{equation}
where $g =\mu_{ge} \sqrt{\omega_p / 2\hbar \epsilon_0 \mathcal{V}}$ defines the atom-field coupling strength between probe beam at frequency $\omega_p$ in quantisation volume $\mathcal{V}$ containg $N$ atoms\ucite{fleischhauer2000darkstate}. The polariton's group velocity is given by $v_g(t) = c \cos^2 \theta(t)$\ucite{fleischhauer2000darkstate}. When the coupling field is strong $\Omega_c\gg g\sqrt{N},\theta\to 0$, then $v_g\to c$ and the polariton is dominated by the photonic component $\langle\hat E\rangle$. When the coupling strength is reduced to zero $\Omega_c=0,\theta=\pi/2$, then $v_g\to 0$, the polariton becomes a purely atomic coherence $\sigma_{gr}$. 

Microscopically, this stored excitation manifests as a spin wave distributed across the atomic ensemble. The collective excitation is described by the $W$-state:
\begin{equation}
|W\rangle = \frac{1}{\sqrt{N}} \sum_{j=1}^N e^{i \mathbf{k} \cdot \mathbf{r}_j} |g_1 \dots r_j\dots g_N\rangle,
\end{equation}
where $\mathbf{k}$ is the excitation laser wave vector and $\mathbf{r}_j$ is the position of the $j$-th atom. This phase factor $e^{i\mathbf{k}\cdot\mathbf{r}_j}$ determines the spatial distribution of the collective atomic coherence, effectively mapping the momentum of the probe photon onto the ensemble. After a controllable storage time,
the stored polariton is converted back into a photon by recovering the control field $\Omega_c$. Due to the phase-matching condition, the collective emission from all atoms interferes constructively only in the direction defined by the original wave vector $\mathbf{k}$. This ensures that the retrieved photon retains both the spatial mode and the quantum coherence of the input field\ucite{distante2017storing,jiao2020singlephoton,jiao2025suppression}. By leveraging the long coherence times of Rydberg states, this storage protocol serves as a versatile interface for quantum state mapping and the synchronization of quantum nodes within a network\ucite{distante2016storage,schmidt-eberle2020darktime}.

\section{Rydberg Nonlinear Quantum Optics}\label{sec:3}

\subsection{Single-photon engineering}
\subsubsection{Single-photon source}

High-efficiency, high-purity, and indistinguishable controlled single-photon sources hold broad application prospects in fields such as quantum computing\ucite{maring2024versatile,obrien2007optical,couteau2023applications}, quantum simulation\ucite{aspuru-guzik2012photonic,hartmann2016quantum}, and quantum secure communication\ucite{hu2016experimental,li2020quantum,qi2019implementation}. The simplest and most intuitive approach is to attenuate a laser pulse to the single-photon level. However, this method produces a weak coherent state rather than a true single-photon source and therefore lacks deterministic controllability. Currently, a wide variety of single-photon sources have been demonstrated, ranging from probabilistically spontaneous parametric down-conversion (SPDC)\ucite{pan2012multiphoton,guo2023ultrathin,kaneda2016heralded} to near-deterministic generation based on quantum dot\ucite{zwanenburg2013silicon,lu2021quantumdot, lodahl2015interfacing,senellart2017highperformance} and NV centers\ucite{zhou2014quantum,song2019generation, ruf2021quantum}. 

In recent years, single-photon sources based on Rydberg atom systems have attracted significant attention owing to their unique advantages. Rydberg atoms exhibit extremely strong dipole-dipole interactions\ucite{gallagher2008dipole,browaeys2016experimental,giudici2025fasta,deleseleuc2017optical}. When there are multiple Rydberg excitations in one atomic ensemble, a strong erasing process of the quantum nature for multiple-photon polaritons proceeds, resulting in nearly complete dephasing on the nanosecond scale as theoretically examined in Refs.~\cite{PhysRevA.85.033811,PhysRevA.86.041802,Bariani2012}. As a consequence, even if there are still multiple Rydberg excitations in the ensemble, there is no quantum coherence between them, thereby annulling the collective effect of the photon retrieval. On the other hand, the single-photon component of the polariton still attains the original quantum coherence inherited during the loading, ensuring that only one coherent collective Rydberg excitation exists within a given region, providing the physical foundation for deterministic single-photon emission. 

\begin{figure}[htbp]
\centering  
\includegraphics[width=0.99\linewidth]{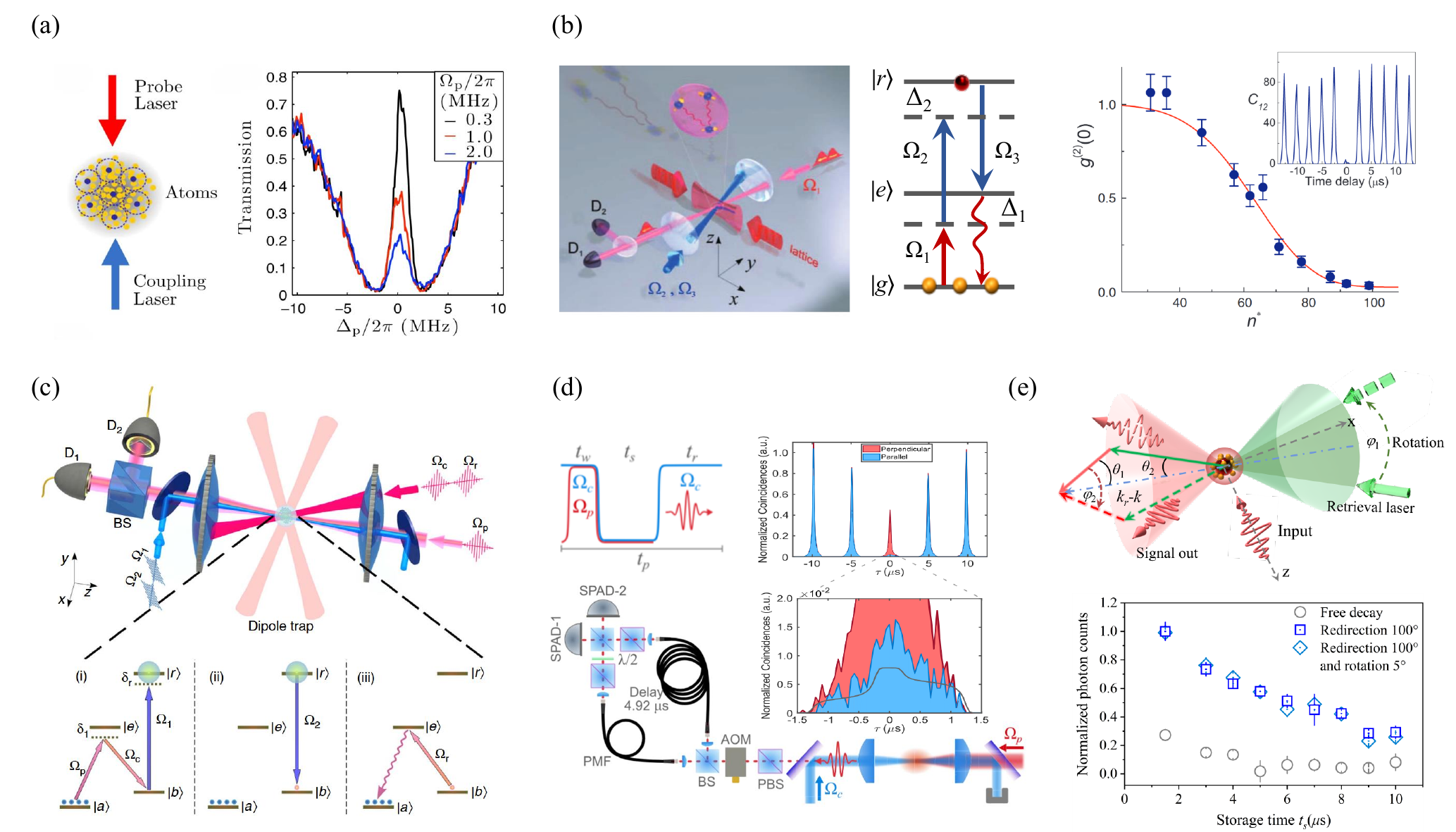}
\caption{Generation of single-photon sources. (a) Rydberg EIT experimental setup and EIT spectra with different probe intensities (reproduced with permission from Ref.~\cite{pritchard2010cooperative}). (b) Experimental scheme and energy level diagram for the generation of single photons within a cold atomic ensemble. And the dependence of the second-order correlation function $g^{(2)}(0)$ on the principal quantum number $n$ (reproduced with permission from Ref.~\cite{dudin2012Strongly}). (c) Experiment setup, level diagram, and experimental protocol for a quantum memory system 
(reproduced with permission from Ref.~\cite{li2016Quantuma}). (d) Realization of an on-demand, indistinguishable single photon source (reproduced with permission from Ref.~\cite{ornelas-huerta2020Ondemand}). (e) Direction-switchable single-photon emitter using a Rydberg polariton (reproduced with permission from Ref.~\cite{li2026}).}
\label{Fig1}
\end{figure}

In 2010, Pritchard \textit{et al.} experimentally demonstrated strong cooperative optical nonlinearities in a Rydberg blockade ensemble using a ladder-type EIT system\ucite{pritchard2010cooperative}. As shown in Fig.~\ref{Fig1}(a), the two-photon resonance transmission decreases with increasing probe intensity, indicating the emergence of strong nonlinear dissipative effects. This cooperative behavior originates from long-range dipole-dipole interactions between Rydberg atoms, which enable strong correlations at the single-excitation level. 

Dudin and Kuzmich demonstrated the first experimental realization of a deterministic single-photon source in 2012\ucite{dudin2012Strongly}. Two nearly collinear counter-propagating laser beams at $795~\rm nm$ and $475~\rm nm$ were used for two-photon excitation, coherently creating collective atomic spin waves between the ground state and the Rydberg state, as shown in Fig.~\ref{Fig1}(b). After a controllable storage time, a $475~\rm nm$ readout laser resonant with the Rydberg state was activated to reconvert the spin wave into an optical field. The emitted photons are incident on a Hanbury Brown-Twiss setup to measure the second-order intensity correlation function $g^{(2)}(0)$. The purity of the output single-photon significantly increases, particularly at $n=102$; the $g^{(2)}(0)= 0.040(14)$ was measured, demonstrating a pronounced photon anti-bunching effect. In 2016, Li and Kuzmich demonstrated a quantum memory system that increases the storage time by almost two orders of magnitude by trapping nonclassical polariton states in the ground atomic energy level\ucite{li2016Quantuma}, shown in Fig.~\ref{Fig1}(c). At principal quantum number $n = 70$, they obtained a highly pure single-photon source with $g^{(2)}(0)$ approaching zero. In 2020, an on-demand indistinguishable single-photon source was demonstrated based on a strongly interacting Rydberg system, shown in Fig.~\ref{Fig1}(d), producing single photons with $g^{(2)}(0) = 5.0(1.6) \times 10^{−4}$ and indistinguishability of 0.980(7)\ucite{ornelas-huerta2020Ondemand}. In Table \ref{tab:single_photon_g2}, we summarize the values of second-order correlation functions in some Rydberg single-photon engineering to demonstrate the purity of the achieved single-photon source.

Recently, our group demonstrated a direction-switchable single-photon emitter using a Rydberg polariton\ucite{li2026}. The Rydberg component of the stored photon is changed using a stimulated Raman transition with a specific intermediate state. By adjusting the direction of the retrieval laser, we can redirect the emitted photon into a rich variety of alternative modes. We experimentally demonstrate a redirection angle of $\sim100^\circ$. Building upon this scheme, we propose a quantum routing of single photons with \textit{N} output channels by rotation of the retrieval laser, where all directions have identical routing efficiency, shown in Fig.~\ref{Fig1}(e). In addition, the protocol suppresses motional dephasing through a Raman $\pi$ pulse that compensates the velocity-dependent phase accumulated by the collective Rydberg excitation. 
As a result, the stored-photon lifetime is extended to $>10~\mu$s ($>20$ times the photon-processing time), enabling functional quantum devices based on Rydberg polaritons.

\begin{table}[htp!]
\centering
\begin{tabular}{ccc}
\hline
\textbf{Year} & \textbf{$g^{(2)}(0)$} & \textbf{Reference} \\
\hline
2012 & $0.040(14)$
     & \textit{Science}\ucite{dudin2012Strongly} \\

2013 & $0.68(4)$
     & \textit{Phys. Rev. Lett.}\ucite{Maxwell2013} \\

2014 & $0.32(18)$
     & \textit{Phys. Rev. A}\ucite{maxwell2014microwave} \\

2016 & $0.00(4)$
     & \textit{Nat. Commun.}\ucite{li2016Quantuma} \\

2020 & $5.0(1.6)\times10^{-4}$
     & \textit{Optica}\ucite{ornelas-huerta2020Ondemand} \\

2021 & $0.42(2)$
     & \textit{Phys. Rev. Lett.}\ucite{spong2021collectively} \\

2023 & $0.027$
     & \textit{Nat. Photon.}\ucite{magro2023deterministic} \\

2024 & $5(5)\times10^{-3}$
     & \textit{Rep. Prog. Phys.}\ucite{xu2024continuously} \\

2026 & $0.34(8)$
     & \textit{Optica}\ucite{li2026} \\
\hline
\end{tabular}
\caption{Second-order correlation functions in Rydberg single-photon engineering.}
\label{tab:single_photon_g2}
\vspace{1mm}
\end{table}

While deterministic high-purity and indistinguishable single photons are achieved based on the Rydberg system, the photon production efficiency is low in free space.  To overcome this challenge, a scheme was demonstrated that maps the internal states of cavity-confined Rydberg superatoms onto optical qubits\ucite{magro2023deterministic}, achieving a single-photon generation efficiency of $60\%$. These results highlight the potential applications of single-photon sources based on Rydberg blockade in QIP. In addition, a theoretically efficient single-photon generation scheme in free space was proposed, in which the process is controlled by a single source Rydberg atom without requiring full confinement of the entire atomic ensemble\ucite{petrosyan2018deterministic}, although it has not been demonstrated
experimentally.

\subsubsection{Coherent manipulation of a single photon using microwave}

Beyond the generation of single photons, coherent manipulation of single-photon states represents another essential capability for QIP. The key challenge in QIP lies in achieving coherent, fast, and programmable manipulation of the quantum state of individual photons. Compared to traditional optical methods, Rydberg atoms exhibit exceptional sensitivity to microwave and radio-frequency external fields due to their large electric dipole moment and highly tunable energy level structure. This unique advantage enables the integration of microwave quantum information interfaces into optical single-photon platforms, providing a powerful tool for coherent manipulation of stored Rydberg excitations using a microwave field. 

\begin{figure}[htbp]
\centering  
\includegraphics[width=0.99\linewidth]{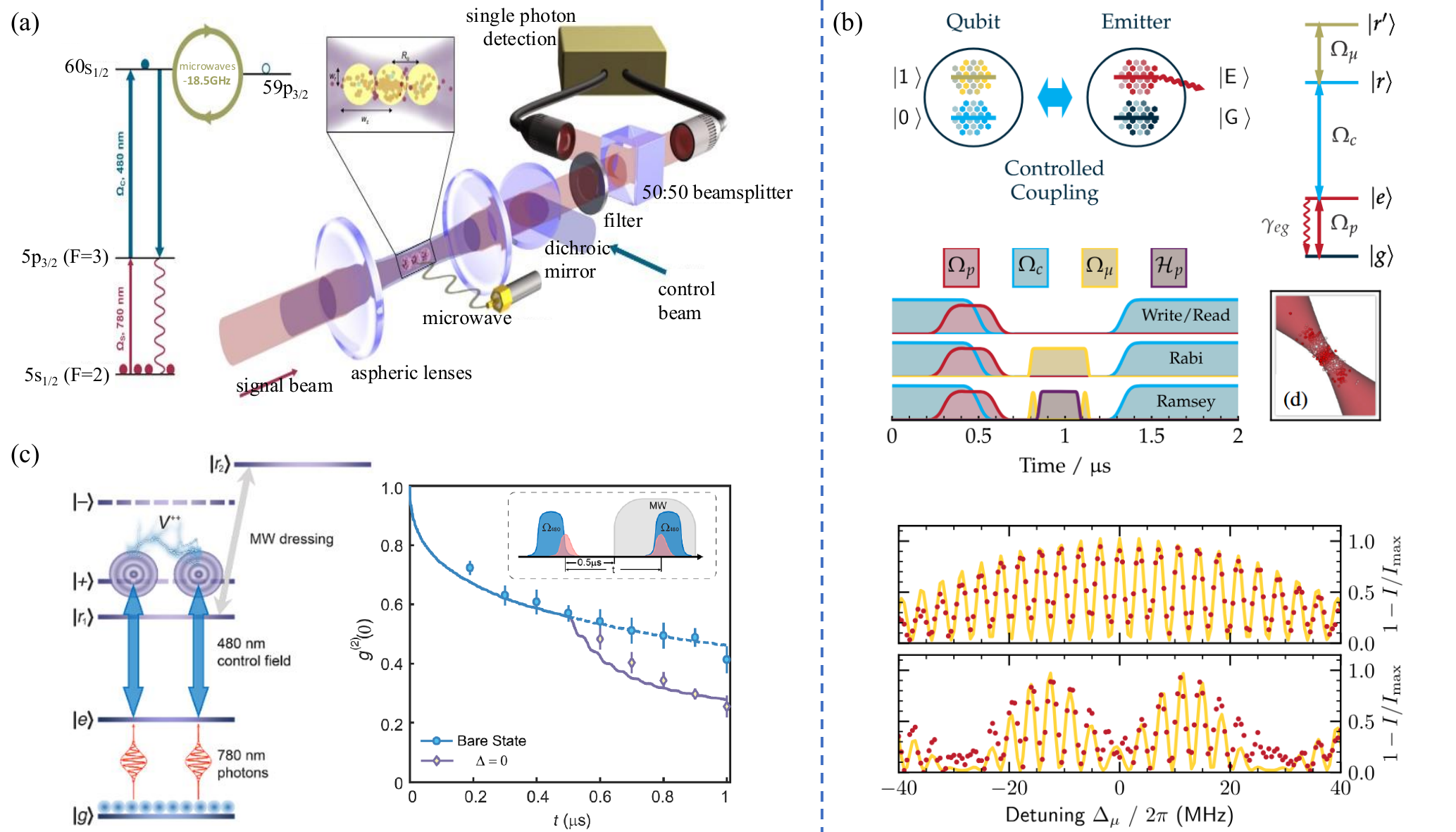}
\caption{Interface between the optical and microwave photons via Rydberg atoms. (a) Schematic of the level scheme and experimental setup for storing optical photons as Rydberg polaritons, with state manipulation facilitated by an external microwave field (reproduced with permission from Ref.~\cite{Maxwell2013}). (b) Experimental scheme for collectively encoded Rydberg qubits and their coherent control. High-fidelity qubit rotations are driven by resonant microwave pulses to realize Ramsey interferometry and the Hadamard gate (reproduced with permission from Ref.~\cite{spong2021collectively}). (c) Microwave-assisted Rydberg state dressing enabling continuously tunable single-photon nonlinearity (reproduced with permission from Ref.~\cite{xu2024continuously}).}
\label{Fig2}
\end{figure}

In 2013 and 2014, Maxwell \textit{et al.} demonstrated microwave field control of quantum states stored in optical photons within a cold Rydberg cloud, enabling rapid quantum bit rotation and control of photon interactions\ucite{Maxwell2013, maxwell2014microwave}, as shown in Fig.~\ref{Fig2}(a). By coupling stored Rydberg polaritons to adjacent high-energy excited states, they observed collective many-body Rabi oscillations, which are highly sensitive to the ratio of microwave drive intensity to resonant dipole-dipole interactions. When the microwave Rabi frequency is varied, the polaritons can smoothly transition from an interaction-dominated decoherence region to a strongly driven region, where their coherent spatial correlations are restored. This ability to dynamically control the interactions between neighboring polaritons provides a powerful interface between the microwave and optical frequency bands. Following this, a collectively encoded Rydberg qubit was demonstrated\ucite{spong2021collectively}. Using a microwave field to drive transitions between highly excited Rydberg states enables rapid coherent control of the encoded qubit. The robustness of the encoded qubits against external perturbations was tested, showing that even after irreversible loss of some atoms from the polariton mode and exposure to environmental noise, the coherence of the collectively encoded qubits was still exceptionally well preserved. In 2023, Fan \textit{et al.} demonstrated on-demand Rabi oscillation and modulation of single photons using microwave, achieving coherent microwave manipulation of single photons based on a Rydberg atomic ensemble\ucite{fan2023manipulation}. In addition, they demonstrate a robust single-photon Ramsey interferometer based on individual Rydberg excitations, where photons were stored as Rydberg polaritons within the atomic ensemble. Due to the collective excitations, the Ramsey interferometer is robust against fluctuations in the number of incident photons and the OD of the atomic ensemble\ucite{fan2023robust}. 

In addition, through the use of microwave-assisted wavefunction engineering, Xu \textit{et al.} demonstrated the capability for continuously tunable single-photon level nonlinearity, enabled by precise control of Rydberg interaction over two orders of magnitude\ucite{xu2024continuously}. This dressing protocol significantly accelerates quantum operations compared to Rydberg dephasing-based mechanisms. By enabling continuous control over interaction strength, the scheme facilitates the rapid preparation of high-quality single photons even at low principal quantum numbers ($n$), as illustrated in Fig.~\ref{Fig2}(c). Unlike blockade protocols that rely on high-$n$ states, this low-$n$ decoherence scheme is less susceptible to additional decoherence and loss, offering a robust pathway for accelerating quantum operations and scaling photonic QIP.

\subsubsection{Single-photon switch and transistor}

\begin{figure}[htp!]
\centering  
\includegraphics[width=0.99\linewidth]{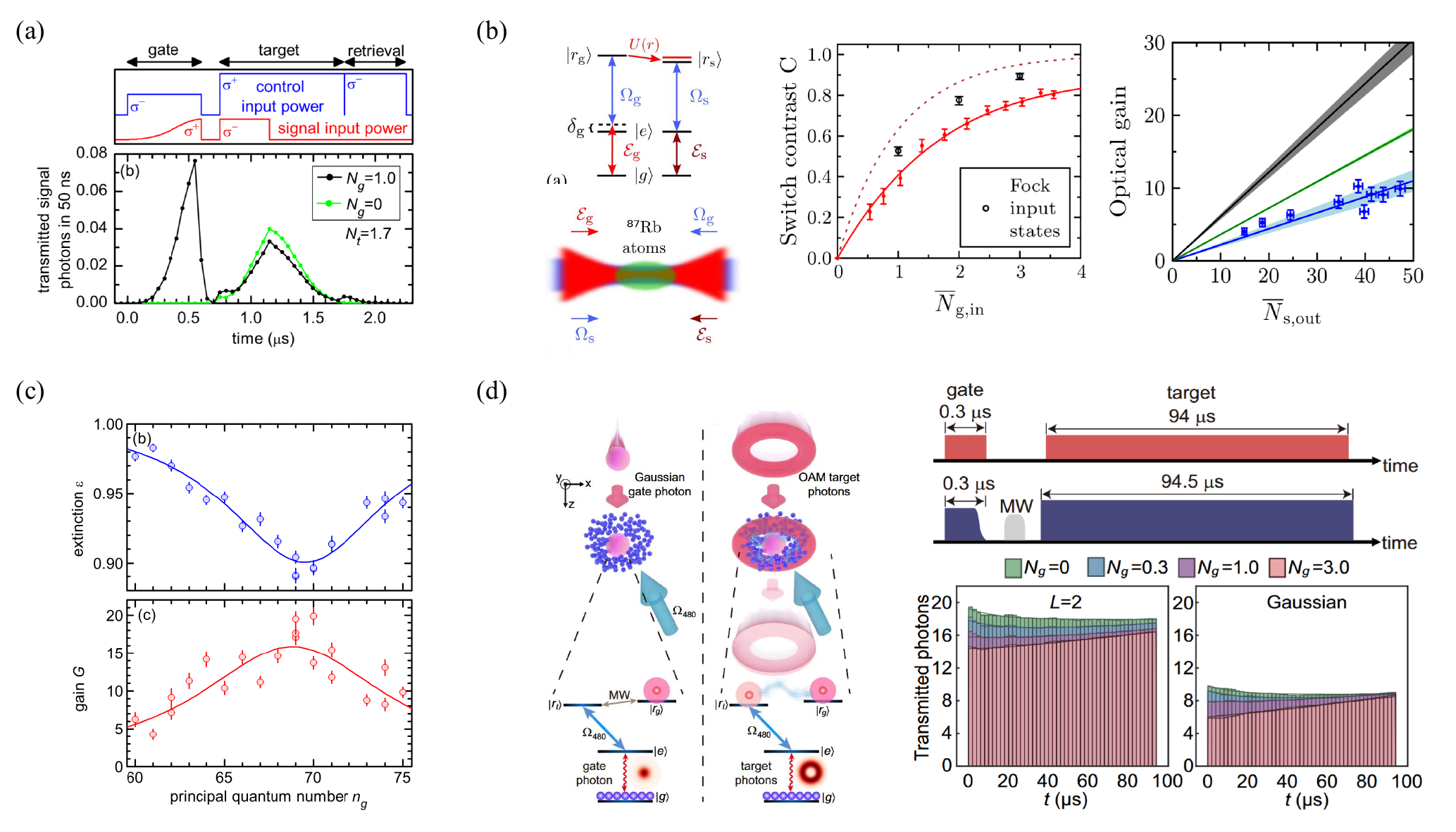}
\caption{All-optical single-photon switches and transistors. (a) Demonstration of the single-photon switch, showing the reduction in transmitted target photons due to a stored gate excitation (reproduced with permission from Ref.~\cite{baur2014SinglePhotonb}). (b) Schematic of the level scheme and setup for the single-photon transistor, together with measurements of the switching contrast and optical gain (reproduced with permission from Ref.~\cite{gorniaczyk2014SinglePhoton}). (c) Optimizing transistors by selecting the most suitable Rydberg state to generate F{\"o}rster resonance (reproduced with permission from Ref.~\cite{tiarks2014SinglePhoton}). (d) Demonstration of a non-local single-photon transistor. Stored Gaussian-gate photons control the transmission of source OAM photons via Rydberg interactions, achieving a significant optical gain enhancement of $G=151$ by reducing self-blocking effects (reproduced with permission from Ref.~\cite{liao2025nonlocal}).}
\label{Fig3}
\end{figure}

All-optical EIT in Rydberg media enables the mapping of long-range interactions between Rydberg atoms onto photons, thereby inducing effective photon–photon interactions. This mechanism serves as the foundation for single-photon switching and transistor functionalities, which are fundamental components for constructing scalable quantum information networks. The first experimental demonstration of a Rydberg-based single-photon optical switch was achieved by Baur \textit{et al.} in 2014\ucite{baur2014SinglePhotonb}, where a gate photon pulse was stored in an ultracold Rydberg gas via the slow-light effect, establishing a blockade volume of radius $R_b$ within the atomic medium. This stored gate excitation exerts a strong Rydberg-Rydberg interaction that shifts the Rydberg level of nearby target atoms out of resonance, effectively collapsing the three-level EIT system into a dissipative two-level system for the subsequently incident target photons\ucite{gorshkov2011photonphoton}. As a result, the target pulse undergoes significant scattering within the blockade sphere, leading to the pronounced suppression of transmission shown in Fig.~\ref{Fig3}(a). In contrast, the target pulse passes with negligible absorption in the absence of a gate photon, thereby realizing a high-contrast all-optical switch. Such Rydberg-based switches offer intriguing prospects for deterministic quantum logic operations, quantum computing, and the development of non-destructive photon detection\ucite{saffman2010quantum}.

Building upon the single-photon switch, the Rydberg photonic transistor introduces the capability of optical amplification, where a single gate photon controls the transmission of a multi-photon source pulse. The performance of such a transistor is quantified by the optical gain, defined as the average number of source photons blocked per incident gate photon: $G = (\bar{N}^{\text{no gate}}_{s,\text{out}} - \bar{N}^{\text{with gate}}_{s,\text{out}})/\bar{N}_{g,\text{in}}$, where $\bar{N}^{\text{no gate}}_{s,\text{out}}$ and $\bar{N}^{\text{with gate}}_{s,\text{out}}$ denote the mean number of transmitted source photons in the absence and presence of a gate excitation, respectively, and $\bar{N}_{g, \text{in}}$ is the mean number of incident gate photons\ucite{tiarks2014SinglePhoton,gorniaczyk2016enhancement}. A gain $G > 1$ signifies that a single gate photon can actively control the flow of multiple source photons, a direct analogue to current amplification in electronic transistors.

This concept was first experimentally demonstrated by Gorniaczyk \textit{et al.}\ucite{gorniaczyk2014SinglePhoton} and Tiarks \textit{et al.}\ucite{tiarks2014SinglePhoton}. When the optical depth within the blockade volume $\text{OD}_b$ is sufficiently large, a high switching contrast $C = 1 - (\bar{N}^{\text{with gate}}_{s,\text{out}} / \bar{N}^{\text{no gate}}_{s,\text{out}}) > 0.9$ can be achieved. As shown in Fig.~\ref{Fig3}(b), the contrast $C$ increases with the gate photon number and eventually approaches the fundamental limit set by the Poisson statistics of the gate pulse. The observed gain, typically ranging from $20$ to $30$, is primarily limited by the self-blockade effect among source photons, which causes the transmission to saturate as the source intensity increases. To mitigate these limitations, Tiarks \textit{et al.} optimized the transistor by selecting specific principal quantum numbers ($n_g=69, n_s=67$) to exploit a near-zero energy mismatch, inducing a F{\"o}rster resonance that enhanced gate-source interactions\ucite{tiarks2014SinglePhoton}. Lower principal quantum numbers not only reduce self-blockade and dephasing but also increase Rydberg state population lifetime at certain experimental densities due to reduced inelastic collision rates. This ultimately achieved a single-photon transistor with an input gate photon count of $1$ and a gain of $G=20$, as shown in Fig.~\ref{Fig3}(c). Building on this, Gorniaczyk \textit{et al.} applied external vacuum electrodes to achieve Stark-tuned F{\"o}rster resonance\ucite{gorniaczyk2016enhancement}, precisely tuning adjacent Rydberg state pairs to resonance and elevating the maximum gain to $G=200$.

A persistent challenge in further enhancing gain is the trade-off between strong gate-source interactions and the detrimental self-blockade among source photons in traditional Gaussian-mode configurations. Recently, Liao \textit{et al.} demonstrated a novel approach utilizing orbital angular momentum (OAM) photons coupled with Rydberg ensembles\ucite{liao2025nonlocal}, as illustrated in Fig.~\ref{Fig3}(d). By exploiting the unique spatial distribution of OAM modes, they successfully suppressed the self-blockade among source photons, achieving high-performance non-local transistors with gains up to $G=151(3)$—a threefold improvement over Gaussian modes under similar conditions. These results demonstrate robust nonlocal control over single photons in OAM modes, opening new avenues for novel quantum devices and multi-photon quantum optics utilizing Rydberg atoms. Table~\ref{gain} summarizes the gains of some Rydberg-based single-photon transistors above.
\begin{table}[htp!]
\centering
\renewcommand{\arraystretch}{1} 
\setlength{\tabcolsep}{4mm} 
\begin{tabular}{ccc}
\toprule
\textbf{Year} & \textbf{Gain($G$)} & \textbf{Reference} \\
\midrule
2014 & 10   & \textit{Phys. Rev. Lett.} \ucite{gorniaczyk2014SinglePhoton} \\
2014 & 20   & \textit{Phys. Rev. Lett.}\ucite{tiarks2014SinglePhoton} \\
2016 & 200  & \textit{Nat. Commun.} \ucite{gorniaczyk2016enhancement} \\
2021 & 17   & \textit{Phys. Rev. Lett.}  \ucite{xu2021fast} \\
2025 & 151  & \textit{Phys. Rev. Lett.} \ucite{liao2025nonlocal} \\
\bottomrule
\end{tabular}
\caption{Summary of Gain ($G$) in Rydberg-based photon transistors.}
\label{gain}
\end{table}

Several other novel approaches to realize single-photon nonlinearities by mapping strong interactions between Rydberg atoms onto traveling photons. For instance, interspecies Rydberg interactions within dual-species atomic mixtures can be used to tune switching responses\ucite{chen2021twocolor}, while combining the Rydberg blockade with precision optical cavities has pushed the switching capacity beyond one thousand photons\ucite{hao2019singlephoton}. Furthermore, the introduction of facilitation-induced transparency (FIT) in spatially separated dual-channel configurations provides a means to preserve the coherence of control atoms more effectively\ucite{ding2023facilitationinduced}. Collectively, these advancements in Rydberg-based single-photon sources, switches, and transistors establish a comprehensive toolbox for single-photon engineering, providing a robust foundation for the realization of scalable quantum information architectures.

\subsection{Quantum photonic gate}

Photonic qubits have emerged as highly promising carriers in quantum information science due to their unique ability to integrate transmission with processing. Especially when supported by fiber-optic networks, photons serve as a natural interconnection medium for future quantum internet and distributed quantum computing\ucite{kimble2008quantum,wehner2018quantum,azuma2023quantum}. In this context, two-qubit operations, such as controlled-NOT (CNOT) and controlled-phase gates, are particularly crucial for constructing a universal set of quantum logic gates\ucite{lloyd1995almost,sleator1995realizable,monroe1995demonstration,bravyi2005universal}. 

Over the past decades, traditional implementations of photonic CNOT gates have primarily followed two categories. The first encompasses schemes based on linear optical quantum computing (LOQC), which induce effective nonlinearity through linear optical elements, post-selection, and measurement. This approach was first demonstrated in a probabilistic two-photon CNOT gate experimentally\ucite{obrien2003demonstration}. While LOQC can theoretically achieve scalable quantum computing via the KLM protocol\ucite{knill2001scheme}, it demands enormous resources and has a low success probability. In early schemes, the theoretical probability limit was only $1/9$, and it decreases exponentially with the number of gates\ucite{kieling2010photonic}. Although probabilistic issues can in principle be addressed through error correction\ucite{franson2002highfidelity}, the fidelity of linear optical quantum logic gates remains constrained by the imperfections of single photons\ucite{aharonovich2016solidstate, knill2001scheme}. Integrated photonics\ucite{pelucchi2022potential,wang2020integrated} holds promise for enhancing the feasibility and scalability of LOQC, but efficiency and resource consumption issues remain significant challenges. An alternative approach involves directly utilizing optical nonlinear media to achieve photon-photon interactions. Early concepts relied on the Kerr effect\ucite{gea-banacloche2010impossibility,imoto1985quantum}, where the electric field of one photon alters the refractive index of a crystal, inducing phase shifts in another photon. However, conventional optical materials exhibit extremely weak nonlinearity at the single-photon level, incapable of generating sufficient conditional phase shifts for single photons\ucite{matsuda2009observation,fushman2008controlled,turchette1995measurement}, such as a $\pi$ conditional phase shift. While such interactions are typically negligible in conventional optical media, Rydberg systems have recently emerged as a powerful platform for engineering strong, controllable interactions between individual photons.

A deterministic photonic phase gate based on Rydberg EIT was theoretically proposed by Friedler \textit{et al.} in 2005\ucite{friedler2005longrange}. Operating in the dispersive regime, this protocol investigates two counter-propagating single-photon pulses within an EIT medium. The dispersive nature of the system arises from interaction-induced energy shifts of Rydberg polaritons, leading to phase accumulation without relying on absorption. In the presence of a static electric field, the Rydberg atoms exhibit large permanent electric dipole moments, giving rise to strong dipole-dipole interactions between pairs of Rydberg atoms. By applying weak-field and adiabatic approximations, they demonstrated that under experimentally realizable conditions, the conditional phase shift accumulated during the collision of two polaritons is spatially homogeneous. Crucially, this homogeneity implies that the requirement for beam focusing in experiments is significantly relaxed. Unlike traditional schemes that necessitate tight focusing to the diffraction limit to achieve sufficient interaction strength, the Rydberg-mediated phase gate can maintain high fidelity even when the transverse cross-section of the photon pulses is considerably larger than the atomic resonant absorption cross-section. This mechanism was subsequently extended to co-propagating configurations\ucite{gorshkov2011photonphoton}, providing a more robust and scalable framework for implementing universal quantum logic gates in neutral-atom platforms.

\begin{figure}[htp!]
\centering  
\includegraphics[width=0.9\linewidth]{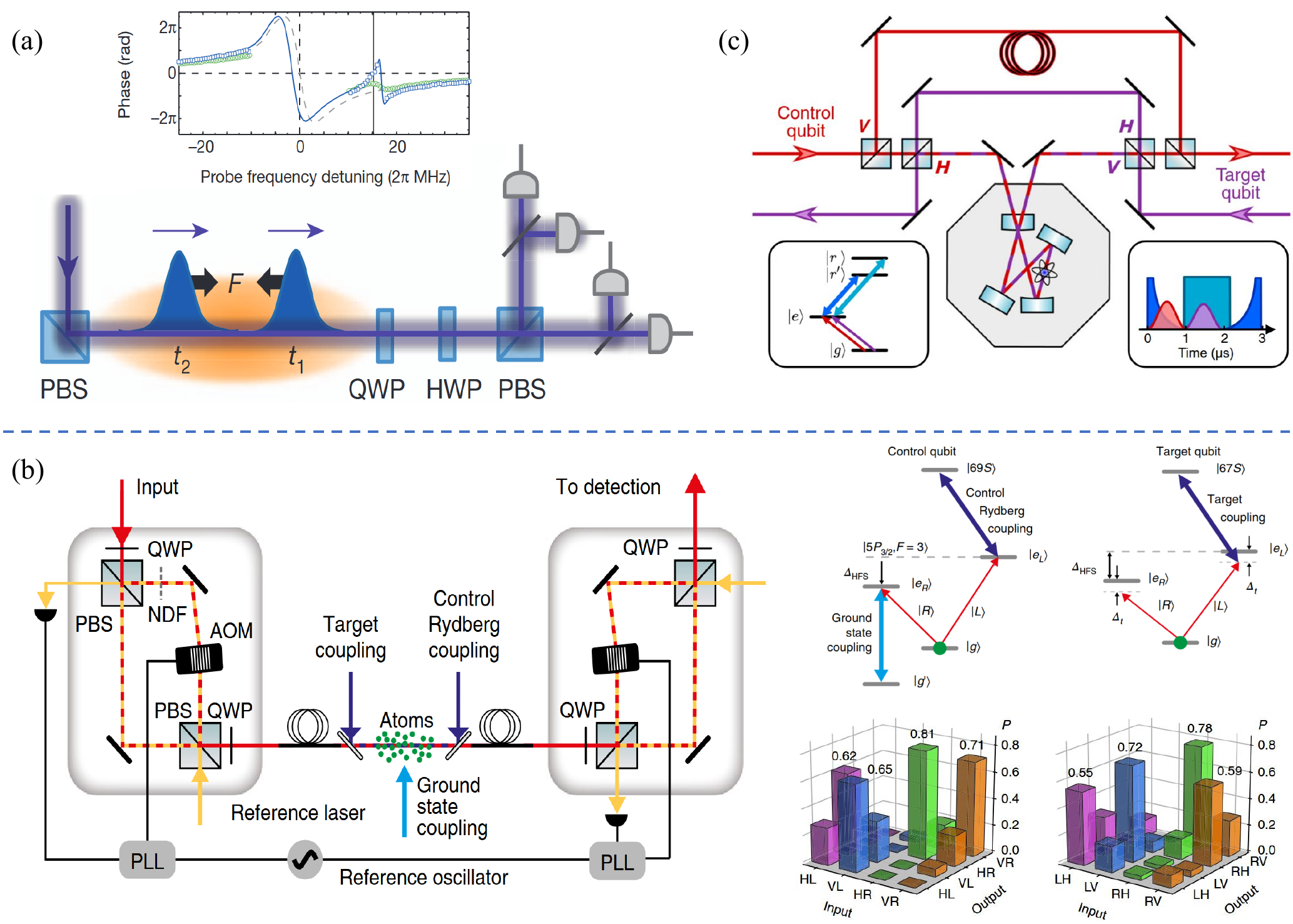}
\caption{(a) Schematic of the experimental setup and the measured nonlinear phase shift of the probe field as a function of frequency detuning (reproduced with permission from Ref.~\cite{firstenberg2013attractive}). (b) Experimental setup and level schemes for the control and target qubits. The scheme illustrates the storage of the control photon's $|L\rangle$ ($|R\rangle$) polarization in the Rydberg state $|69S\rangle$ (ground state $|g'\rangle$), and the propagation of the target photon's $|L\rangle$ component as a Rydberg polariton involving the $|67S_{1/2}\rangle$ state and the $|R\rangle$ component propagates in a non-resonant mode (reproduced with permission from Ref.~\cite{tiarks2019Photon}). (c) Schematic of the CNOT gate implementation showing the dual-rail encoding of polarization qubits and the conditional $\pi$ phase shift mediated by cavity-enhanced Rydberg EIT (reproduced with permission from Ref.~\cite{stolz2022quantumlogic}).}
\label{Fig4}
\end{figure}

The realization of practical quantum gates relies on the experimental realization of significant dispersive nonlinearity. Unlike absorptive interactions used in optical switches, dispersive regimes allow photons to acquire substantial conditional phase shifts while minimizing dissipative loss. A key approach involves detuning the signal and control fields from their respective Rydberg excitation levels to suppress absorption. Within an optical cavity, such detuning enables the observation of enhanced $\chi^{(3)}$ dispersive effects in cold Rydberg ensembles\ucite{parigi2012observation}. This setup leverages the cavity's interferometric sensitivity to map single-photon nonlinearities onto measurable shifts in the cavity resonance frequency. Beyond cavity-based measurements, strong photon-photon interactions have been directly observed in free-space Rydberg media. As shown in Fig.~\ref{Fig4}(a), time-resolved quantum state tomography was employed to observe two-photon bound states formed by strong attractive interactions between Rydberg polaritons\ucite{firstenberg2013attractive}. The observation of such bound states provides direct evidence of strong effective photon-photon interactions and demonstrates the ability of the system to generate large conditional phase shifts exceeding one radian, approaching the regime required for a $\pi$ phase gate. Furthermore, larger nonlinear phases and the emergence of triphoton bound states have also been observed\ucite{liang2018observation}, confirming the scalability of these strong dispersive interactions. Collectively, these results demonstrate that Rydberg-EIT systems provide strong dispersive nonlinearities capable of generating substantial conditional phase shifts, approaching the regime required for deterministic photon-photon phase gates.

While dispersive interactions have demonstrated strong nonlinear phase shifts, deterministic quantum logic requires a mechanism where a single photon directly dictates the propagation of another. This requirement naturally leads to blockade-based protocols. Tiarks \textit{et al.} reported the coupling between Rydberg blockade and EIT in a dense, cold atomic gas with large OD\ucite{tiarks2016optical}. A control pulse averaging $0.6$ photons was stored in the atomic gas, while a second target optical pulse averaging $0.9$ photons propagated through the medium via EIT timing. If the Rydberg excitations of the control photon block the entire medium, the EIT signature of the target pulse shifts to a different frequency. Upon detecting the retrieval photons from the control pulse, a controlled phase shift of $3.3\pm 0.2$ radians, linearly dependent on atomic density, was measured for the target pulse. This experiment demonstrated that a single stored Rydberg excitation can imprint a controllable phase shift on a propagating photon, establishing the feasibility of blockade-based photonic quantum gates.

The first deterministic photon-photon gate based on Rydberg interactions was experimentally demonstrated in the polarization basis [see Fig.~\ref{Fig4}(b)]\ucite{tiarks2019Photon}. In this protocol, different polarization components of a control photon were stored in either a Rydberg state or a ground-state level of an ultracold atomic ensemble. Subsequently, the $|L\rangle$ component of the target photon propagates under Rydberg EIT conditions and interacts with the Rydberg component stored in the control photon. In contrast, the $|R\rangle$ component propagates in a non-resonant mode without experiencing EIT. Finally, both polarization components of the control qubit are retrieved. Data are post-selected upon detection of one photon in each pulse to compensate for non-unity efficiencies. The conditional phase shift originated from the Rydberg-Rydberg interaction during the temporal overlap of the two excitations. Optimization near a F{\"o}rster resonance (between $67S_{1/2}$ and $69S_{1/2}$) was crucial to suppress dephasing, ultimately yielding a CNOT gate fidelity of $F_{\rm CNOT}=70(8)\%$. Despite these milestones, achieving both high efficiency and high fidelity in deterministic photon-photon gates remains a central challenge. A hybrid approach combining near-optimal Rydberg single-photon sources with linear optical protocols has demonstrated photonic CNOT gates with fidelities as high as $99.84(3)\%$\ucite{shi2022Highfidelityb}. Unfortunately, despite the high fidelity, such schemes remain fundamentally constrained by the probabilistic nature of post-selection-based protocols.

Moreover, in cavity-Rydberg atom hybrid systems, it is possible to achieve deterministic photon gates with higher efficiency and fidelity. As shown in the configuration in Fig.~\ref{Fig4}(c), the medium is placed within a ring cavity. Control photons are stored in an atomic ensemble in Rydberg states, and target photons then collide with the atomic ensemble\ucite{stolz2022quantumlogic}. In the absence of control photons, the target photon undergoes a Rydberg EIT process with high transmission. However, if Rydberg excitations from the control photon are present, interactions between the two Rydberg states cause the system to deviate from the two-photon resonant EIT process, leading to reduced transmission of the target photon. Consequently, the target photon undergoes conditional loss in the presence of the control photon, with the conditional $\pi$ phase shift being converted into a photon gate by the cavity and polarization encoding. Compared to free-space schemes, the cavity enhances the coupling strength between light and atoms, enabling the use of lower atomic densities. The reduced atomic density suppresses interatomic collisions, significantly lowering the decoherence rate of Rydberg states. Despite laser phase noise limitations, a two-photon CNOT gate with an average efficiency of $41.7(5)\%$ and a post-selection fidelity of $81(2)\%$ was achieved. This experiment demonstrates the immense potential of combining cavities with Rydberg quantum nonlinearity, providing a realistic technical pathway for logical interconnections between optical QIP and optical quantum network nodes. Table~\ref{photon_gate} summarizes representative Rydberg-mediated photonic quantum gate schemes reported over the past two decades, highlighting their key performance metrics and facilitating a direct comparison between different implementation strategies.




\begin{table*}[htp!]
\centering
\renewcommand{\arraystretch}{1}
\begin{tabular}{c l l l}
\toprule
\textbf{Year} &
\textbf{Reference} &
\textbf{Scheme} &
\textbf{Representative metric} \\
\midrule
2005 &
\textit{Phys. Rev. A} \ucite{friedler2005longrange}
&
Phase gate
&
Deterministic $\pi$ phase gate (proposal)
\\
2013 &
\textit{Nature} \ucite{firstenberg2013attractive}
&
Nonlinear phase
&
Conditional phase shift: $>1~\mathrm{rad}$
\\
2016 &
\textit{Sci. Adv.} \ucite{tiarks2016optical}
&
Controlled phase
&
Conditional phase shift: $3.3\pm0.2~\mathrm{rad}$
\\
2019 &
\textit{Nat. Phys.} \ucite{tiarks2019Photon}
&
Photonic CNOT
&
Gate fidelity: $70(8)\%$
\\
2022 &
\textit{Nat. Commun.} \ucite{shi2022Highfidelityb}
&
Photonic CNOT
&
Gate fidelity: $99.84(3)\%$ (post-selected)
\\

2022 &
\textit{Phys. Rev. X} \ucite{stolz2022quantumlogic}
&
Cavity photonic CNOT
&
Gate fidelity: $81(2)\%$; efficiency: $41.7(5)\%$
\\
\bottomrule
\end{tabular}
\caption{Comparison of representative Rydberg-mediated photonic quantum gate schemes.}
\label{photon_gate}
\end{table*}
In summary, Rydberg-mediated photon-photon quantum gates have advanced along three primary directions: dispersive phase gates, blockade-based conditional gates, and cavity-enhanced hybrid systems. These approaches provide promising pathways to overcome the fundamental limitations of conventional linear optics and weakly nonlinear media, enabling deterministic photon-photon interactions. While conditional phase shifts approaching $\pi$ have been demonstrated, remaining challenges include photon loss, finite optical depth, and decoherence of Rydberg excitations, which limit the achievable efficiency and fidelity. Continued improvements in atomic coherence, optical depth, and cavity coupling strength are expected to further enhance gate performance. Future progress in these directions may ultimately enable scalable photonic quantum logic, efficient photonic Bell-state detection, and integrated quantum network architectures.

\subsection{Contactless nonlinear optics}

\begin{figure}[htp!]
\centering  
\includegraphics[width=0.7\linewidth]{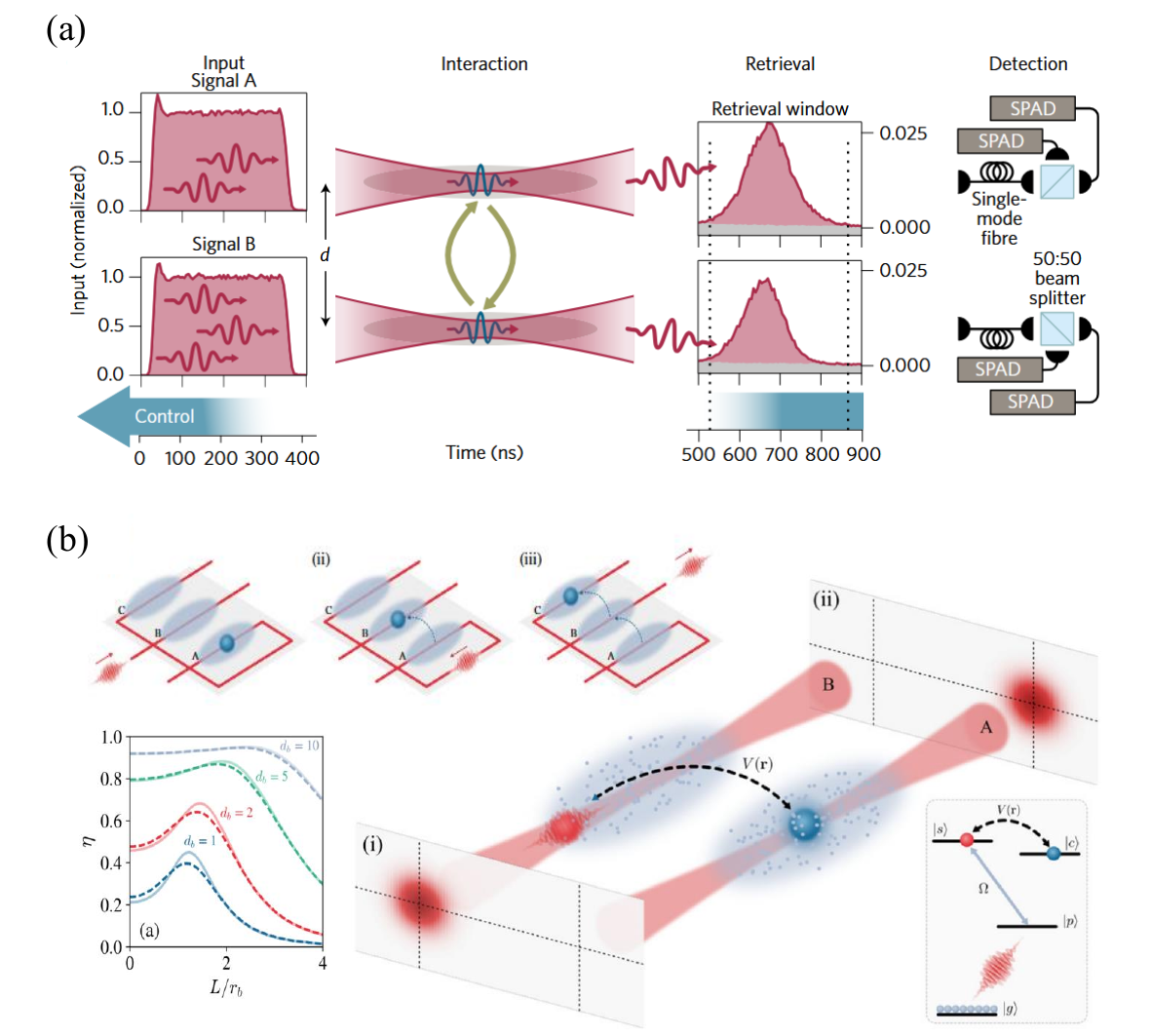}
\caption{(a) Experimental realization of contactless nonlinear interactions between photons stored in spatially separated atomic channels without physical overlap. The measured cross-correlations demonstrate a tunable long-range coupling between the isolated modes that depends on the Rydberg principal quantum number and the separation distance (reproduced with permission from Ref.~\cite{busche2017Contactlessb}). (b) Schematic of the dual-channel setting for dipolar polariton exchange and the optical network with integrated feedback designed to implement a photonic CZ gate (reproduced with permission from Ref.~\cite{khazali2019Polaritonb}).}
\label{Fig8}
\end{figure}

In Rydberg quantum optics, optical photons can be coherently and reversibly mapped onto collective Rydberg excitations, whereby strong dipole–dipole interactions between atoms are transduced into long-range effective photon–photon interactions. This mechanism bypasses the requirement for direct spatial overlap between optical modes, enabling contactless nonlinear optics. The first observation of contactless coupling between photons stored in spatially separated media was reported by Busche \textit{et al.} in 2017\ucite{busche2017Contactlessb}. As shown in Fig.~\ref{Fig8}(a), signal photons are stored as Rydberg collective excitations within two spatially separated atomic ensembles. A shared, wide-waist control laser covers both signal modes, facilitating simultaneous storage and retrieval. Although the ultimate goal for this setup is to engineer a uniform polariton-polariton interaction between the two atomic ensembles, the finite extension of the two atomic ensembles, however, results in an inhomogeneity in the interaction. To understand this, we note that if each of the two atomic ensembles A and B are point-like, then the Rydberg-Rydberg interaction between the two ensembles is certain. But due to the elongated storage region of each photon in the atomic ensembles, the interaction between two Rydberg atoms, one at $\mathbf{r}_{A\alpha}$, the other at $\mathbf{r}_{B\beta}$, depends on the magnitude and orientation of the vector $\mathbf{r}_{A\alpha}-\mathbf{r}_{B\beta}$. This vector, being a function of the two atoms $\alpha$ and $\beta$, leads to a unique interaction for each atom pair $\alpha$ and $\beta$.   

Such type of nonuniform interaction can induce a strong erasing process of the quantum nature of the polaritons. Nonetheless, it also offers a convenient tool for probing the existence of the contactless polariton-polariton interaction.
During storage, the inhomogeneous interaction potential $V(\mathbf{r}_{A\alpha}-\mathbf{r}_{B\beta})$ between Rydberg atoms in adjacent channels induces a temporally accumulated, spatially inhomogeneous phase shift. Upon readout, these phase gradients distort the original photonic modes, manifesting as anti-correlations between the channels. To quantify this cross-channel coupling, the second-order cross-correlation function $g^{(2)}_{AB} = \frac{\langle N_A N_B \rangle}{\langle N_A \rangle \langle N_B \rangle}$ was employed, where $N_A(N_B)$ are the number of photons retrieved in channel A(B). While uncorrelated retrieval yields $g^{(2)}_{AB} = 1$, the experimental result of $g^{(2)}_{AB} = 0.40 \pm 0.03$ (for a separation $d = 10~\mu\text{m}$ and $n = 80$) provides unambiguous evidence of long-range interactions between spatially isolated photons. The interaction strength scales with the principal quantum number $n$ and inversely with the distance $d$, as evidenced by the statistical fluctuations observed in the retrieved pulses. These results demonstrate that non-local control over single photons can be achieved even when their spacing is much larger than the optical wavelength, offering a scalable approach for all-optical QIP. Of course, before this ultimate goal is achieved, the inhomogeneity issue of the accumulated phase should be addressed.

Beyond dispersive effects, contactless Rydberg interactions offer diverse mechanisms for all-optical logic. Recently, a free-space separated contactless single-photon switch based on FIT was proposed\ucite{ding2023facilitationinduced}. In this dual-channel framework, the transparency of a target channel is conditioned on the presence of Rydberg excitations in a remote control channel, where the FIT window width broadens with increasing interaction strength. This remote Rydberg excitation scheme exhibits robust tolerance to experimental parameter fluctuations, thereby preserving atomic coherence. Notably, this mechanism may suppress the formation of ultralong-range Rydberg molecules—typically an undesired byproduct of collisions between ground-state and Rydberg atoms. Such contactless paradigms, ranging from dispersive phase accumulation to resonant exchange and FIT, establish a versatile toolkit for designing complex, non-local quantum architectures.

Furthermore, mapping photons onto Rydberg polaritons with dipolar exchange interactions allows them to inherit effective long-range exchange coupling. During their interaction, the polaritons can coherently exchange their internal states while accumulating a symmetry-protected nonlinear phase shift of $\pi/2$\ucite{thompson2017symmetryprotected}. Unlike conventional schemes, the magnitude of this phase shift is inherently robust against experimental noise, as it is determined by the underlying symmetry of the interaction rather than precise pulse timing or intensity. When extended to multi-channel configurations, this exchange mechanism enables the coherent transfer of photonic information between spatially distinct modes, providing a robust platform for constructing nonlinear photonic networks.

Building on this principle, Khazali \textit{et al.} theoretically propose and analyze the photon hopping mechanism between dual-channel structures in 2019\ucite{khazali2019Polaritonb}. As depicted in Fig.~\ref{Fig8}(b), a stationary Rydberg $P$-polariton in one channel undergoes state exchange with a propagating $S$-polariton in the adjacent channel due to dipole interaction between Rydberg $S$ and $P$ states, rather than the dissipative interactions caused by conventional Rydberg blockade. A counter-intuitive finding of their study is that the photon exchange efficiency does not peak at minimum channel separation. Instead, it increases with spacing and reaches its maximum at an optimal distance—a behavior stemming from the competition between the dipole blockade and the exchange process. When the polariton spacing is too close (within the blockade radius $r_b$), the interaction potential exceeds the EIT linewidth, disrupting EIT conditions and causing the interaction to become purely dissipative. However, since the effective hopping radius $r_h$ scales with the optical depth within the blockade radius $\rm OD_b$ via $r_h=\sqrt{\rm OD_b/2}\,r_b$\ucite{thompson2017symmetryprotected}, large OD conditions allow for an optimized geometry where exchange occurs outside the dissipative blockade region. This geometric configuration preserves the coherence of long-range dipole interactions, enabling efficient coherent jumping of photons between spatial channels on a macroscopic scale. Using this exchange mechanism, they optimized a photonic network to realize an efficient symmetry-protected controlled-Z (CZ) quantum gate based on inherently integrated nonlinear optical feedback [Fig.~\ref{Fig8}(b)].

Rydberg-atom-mediated long-range interactions between spatially separated photons open the door to a variety of promising research directions, including interacting many-body systems\ucite{browaeys2020many}, coherent energy transport\ucite{ravets2014coherent}, quantum simulation of spin models\ucite{weimer2010rydberg}, and scalable all-optical QIP based on tweezers arrays\ucite{sumarac2026controlling}. Neutral atoms hold immense potential for large-scale quantum computing, but accurate entanglement among them relies on strong Rydberg interactions, which severely constrain the distance between atoms. Contactless quantum interactions do not necessarily depend on strong blocking conditions. Based on this, Shi proposed a protocol for achieving high-fidelity quantum operations through phase accumulation in the weak-interaction regime\ucite{shi2021quantum}, where the intrinsic precision of the gate is limited only by the decay of Rydberg states within the interaction range. In conclusion, by breaking the geometric constraints of conventional nonlinear optics, contactless Rydberg interactions not only enhance the robustness of quantum gates but also provide the foundational flexibility required for the next generation of modular and integrated photonic quantum networks.

\subsection{Quantum entanglement}

\subsubsection{Atom-photon entanglement}

Atom–photon entanglement plays a central role in quantum networks\ucite{volz2006observation,simon2007singlephoton,ritter2012elementary}, enabling quantum state transfer between stationary atomic qubits and flying photonic qubits. Rydberg atomic ensembles provide a particularly powerful platform for generating such entanglement, owing to strong and controllable interactions. Over the past decade, experimental progress has shifted from probabilistic generation schemes towards deterministic protocols with improved efficiency and fidelity.

\begin{figure}[htp!]
\centering  
\includegraphics[width=0.99\linewidth]{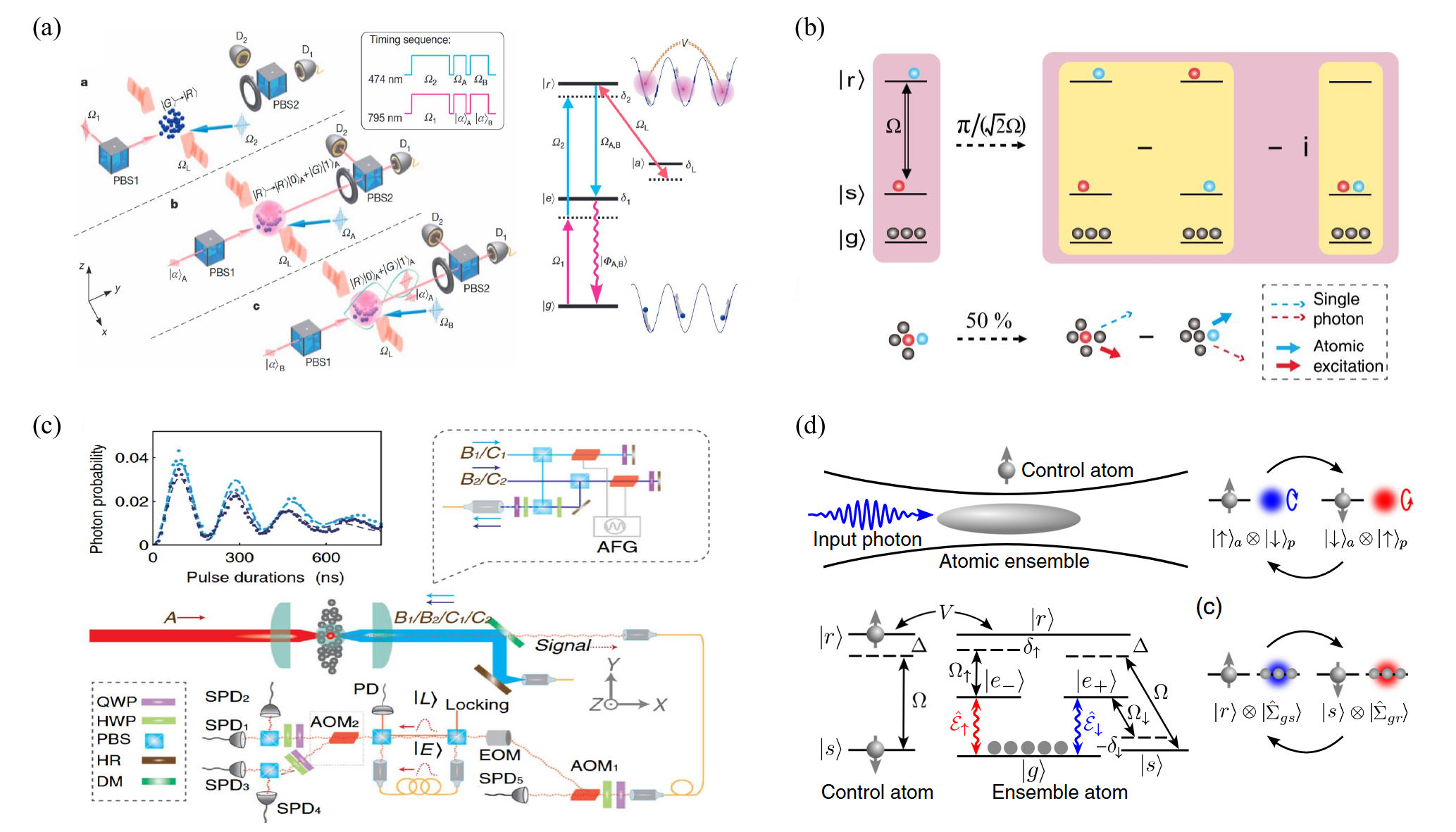}
\caption{(a) An ultracold atomic gas confined in a one-dimensional optical lattice is driven into a singly excited Rydberg state to generate and subsequently map atom-photon entanglement (reproduced with permission from Ref.~\cite{li2013Entanglementa}). (b) Rydberg blockade generates two momentum-distinguishable collective atomic excitations, enabling semi-deterministic atom-photon momentum entanglement by converting one excitation into a single photon (reproduced with permission from Ref.~\cite{li2019semideterministic}). (c) A cold atomic ensemble driven by two-photon Raman transitions generates collective excitations, while an unbalanced Mach-Zehnder interferometer retrieves and measures time-bin entangled single photons (reproduced with permission from Ref.~\cite{sun2022deterministic}). (d) A spin-exchange collision occurs between an input photon and a single control atom, mediated by an atomic ensemble via a Rydberg-dressed energy level structure (reproduced with permission from Ref.~\cite{yang2020atomphoton}). }
\label{Fig9}
\end{figure}

An innovative approach to achieving entanglement between atomic collective excitations and light fields was validated by Li \textit{et al.} in 2013\ucite{li2013Entanglementa}. In this experiment, atoms were confined within a one-dimensional optical lattice, and a specific magic wavelength was employed to trap atoms in both ground and Rydberg states, which suppressed atomic loss while preserving atomic coherence. Through two-photon excitation, the Rydberg blockade effect drove the atomic ensemble from the collective ground state $|G\rangle$ to the singly excited state $|R\rangle$. As shown in Fig.~\ref{Fig9}(a), by applying a readout light field $\Omega_{A}$, the excited atomic component was partially mapped to the light field $|\Phi\rangle_A$, thereby preparing an atom-photon entangled state $|R\rangle |0\rangle_A + |G\rangle |1\rangle_A$. Verification of entanglement relies on phase-sensitive measurements of the optical field, by interfering the readout field with an orthogonally polarized field at a polarization beam splitter and performing coincidence measurements with a single-photon detector. Violation of Bell's inequality validates the entanglement between the atom and the photon. The preparation efficiency of atom-photon entangled states, accounting for factors such as initial state preparation, optical transmission, and detection efficiency, reaches approximately $8\%$. This work established the feasibility of generating atom–photon entanglement via Rydberg blockade and collective excitation mapping.

To enhance the probability of atom-photon entanglement, a semi-deterministic scheme based on momentum degrees of freedom was proposed and experimentally demonstrated\ucite{li2019semideterministic}, as shown in Fig.~\ref{Fig9}(b). The key idea of this work is to encode collective excitations into distinguishable momentum modes, which provide an additional degree of freedom for controlled interference. In this scheme, the atomic ensemble was first prepared into a state containing two collective excitations $|R_2, S_1 \rangle$ with different momenta, corresponding to a ground-state spin excitation $|s\rangle$ and a Rydberg excitation $|r\rangle$. Raman coupling was then used to drive coherent coupling between these excitations.  In the absence of interactions, this would result in independent Rabi oscillations of the two excitations. However, under Rydberg blockade conditions, the double Rydberg-excited state is energetically forbidden, forcing the system to evolve within a restricted Hilbert space. As a result, interference between the two momentum modes leads to the generation of an atom–photon entangled state with a probability of $50\%$, representing a significant improvement over earlier probabilistic schemes and enabling semi-deterministic operation. The entanglement, formed between the polarization of a single photon and the momentum of the remaining atomic excitation, was verified with a fidelity of approximately $0.90$.

The semi-deterministic scheme was further advanced toward deterministic operation by employing time-bin encoding\ucite{sun2022deterministic}, as shown in Fig.~\ref{Fig9}(c). The core principle involves creating a superposition of two collective atomic excitations with distinct temporal modes, thereby enabling controlled photon emission into well-defined early and late time bins. In this scheme, after preparation of a superposition state $(|R_1\rangle + |R_2\rangle)/\sqrt{2}$ in the atomic ensemble, a cyclic sequence of retrieval and re-excitation was applied, allowing the two components to emit photons sequentially while preserving the atom–photon entanglement. As a result, photons were generated deterministically in either early or late time modes, forming a time-bin entangled state between the atomic excitation and the emitted photon. This protocol achieved deterministic generation of atom–photon entanglement with a measured fidelity of $F = 87.8\%$, representing a major step beyond previous semi-deterministic schemes with success probabilities limited to $50\%$.

Beyond blockade-based approaches, mechanisms utilizing Rydberg dressing have been proposed to generate robust atom–photon entanglement\ucite{yang2020atomphoton}, as shown in Fig.~\ref{Fig9}(d). In this theoretical proposal, a single photon exchanges spin states with a single atom through the interaction mediated by an ensemble of Rydberg-dressed atoms. Compared with blockade-based schemes, Rydberg dressing enables long-range interactions while minimizing the population of short-lived Rydberg states, thereby enhancing robustness against decoherence. In the strongly interacting dissipative regime, the system exhibits pronounced loss processes, and dissipation plays a constructive role in driving the system toward a stable eigenstate that is immune to further decay, known as an entangled dark state. As the effective optical depth increases, the system is stabilized into this dark state, enabling robust preparation of atom–photon entanglement through dissipation engineering. This mechanism provides an alternative route to high-fidelity atom–photon entanglement beyond conventional blockade-based protocols.

\subsubsection{Photon-photon entanglement}

Conventional approaches to the generation of photon–photon entanglement based on SPDC and linear optical interference rely on probabilistic photon generation and post-selection, which fundamentally limit their success probability and scalability. In contrast, the strong optical nonlinearities mediated by Rydberg interactions in cold atomic ensembles provide a promising route toward deterministic photon–photon entanglement. 

Ghosh \textit{et al.} proposed a deterministic and heralded high-dimensional hyper-entangled photon generation scheme based on Rydberg cavity QED, providing a promising approach for the development of quantum light sources\ucite{ghosh2021creating}. The central idea of this scheme is to exploit Rydberg blockade together with two-photon spontaneous emission (TPE) processes, utilizing an optical cavity to enhance the desired two-photon emission channel and suppress competing single-photon decay. In this process, the prepared Rydberg atoms undergo cavity-assisted TPE and emit a pair of entangled photons within the telecommunication frequency band. Meanwhile, an additional photon emitted in a subsequent decay process serves as a heralding signal, enabling non-destructive confirmation of the photon-pair generation. Furthermore, by properly selecting the atomic quantization axis relative to the cavity geometry, polarization entanglement can be introduced in addition to energy entanglement, leading to high-dimensional hyper-entangled photon states involving both frequency and polarization degrees of freedom, which offer enhanced channel capacity and improved robustness for quantum communication.

\begin{figure}[htp!]
\centering  
\includegraphics[width=0.99\linewidth]{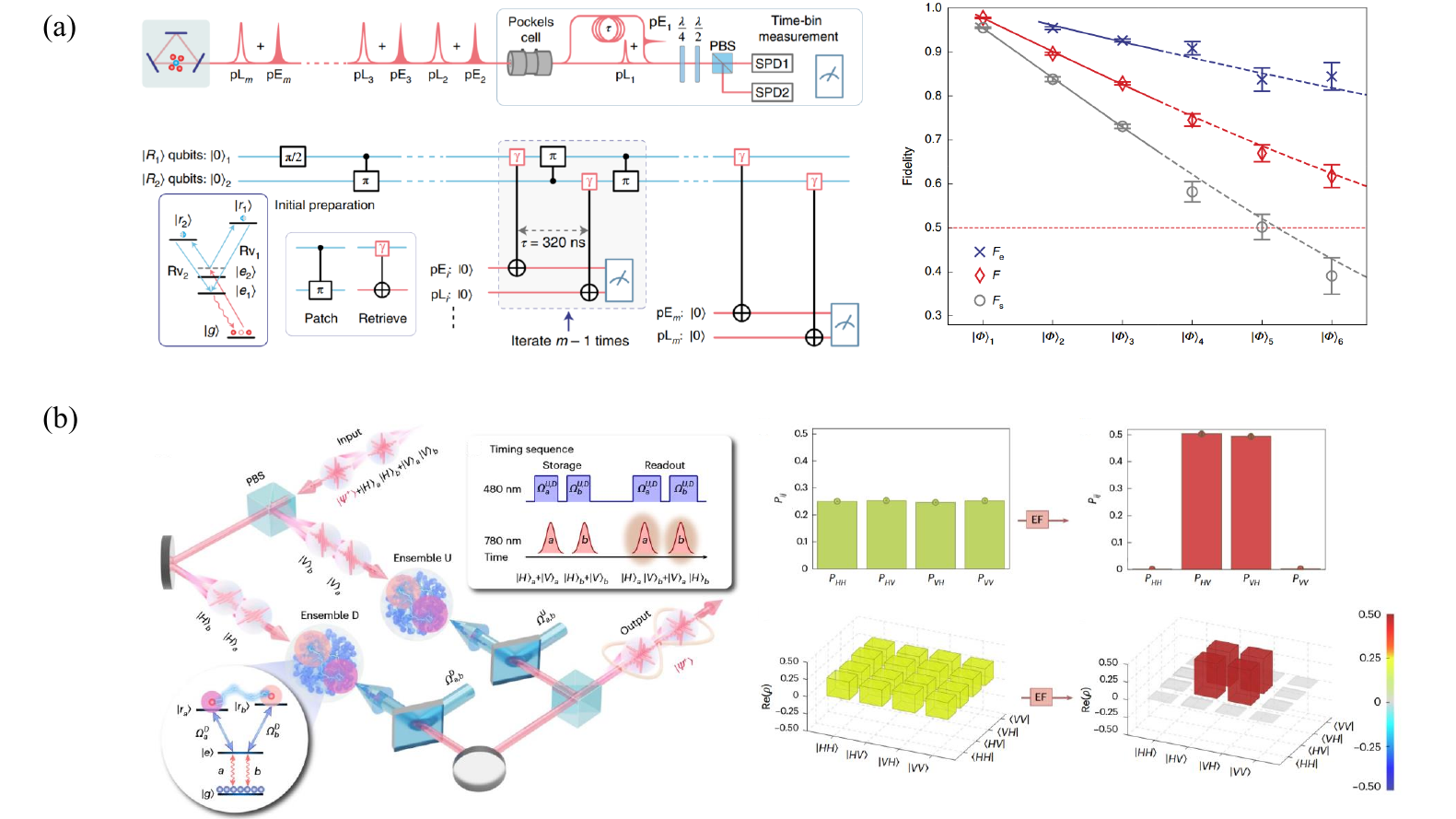}
\caption{(a) A Rydberg superatom sequentially generates time-bin multiphoton entangled states through iterative retrieve-and-patch operations. Measured fidelities confirm the generation of genuine GHZ states for up to six photons, with performance remaining well above the classical threshold (reproduced with permission from Ref.~\cite{yang2022sequential}). (b) Cold Rydberg ensembles mediate deterministic photon-photon interactions, serving as a photonic entanglement filter. The Rydberg blockade effect suppresses undesired noise components, effectively purifying high-fidelity two-photon entanglement from noisy input states (reproduced with permission from Ref.~\cite{ye2023photonic}).}
\label{Fig10}
\end{figure}

As a representative example of sequential multiphoton generation, Yang \textit{et al.} demonstrated the deterministic production of time-bin multiphoton entangled states based on Rydberg blockade\ucite{yang2022sequential}. In this scheme, two different Rydberg states $|r_{1}\rangle$ and $|r_{2}\rangle$ were used to encode qubits in the atomic ensemble, and an initial atomic entangled state $|\Psi\rangle_a = (|0\rangle_1|1\rangle_2 + |1\rangle_1|0\rangle_2)/\sqrt{2}$ was first prepared. Subsequently, through iterative execution of the retrieving and patching operation sequence, the Rydberg collective excitation is first converted into a single photon in the early mode and immediately reconstructed Rydberg excitation. Then a similar operation is performed on $|r_{2}\rangle$ qubit to generate a late mode photon, as shown in Fig.~\ref{Fig10}(a). After iterating the retrieving and patching sequence $m-1$ times, the excitations on $|r_{1}\rangle$ and $|r_{2}\rangle$ can be retrieved to generate $2m$ temporal modes multiphoton entangled states $|\Phi\rangle_m = (|E\rangle^{\otimes m} + |L\rangle^{\otimes m})/\sqrt{2}$. A specially designed annular cavity was employed to enhance collective radiation efficiency and realize an efficient single-photon interface. Experimentally, high-fidelity multiphoton entangled states were achieved, with measured fidelities of $89.6 \pm 0.3\%$, $82.9 \pm 0.3\%$, and $61.8 \pm 2.6\%$ for two-photon Bell states, three-photon, and six-photon Greenberger-Horne-Zeilinger (GHZ) states, respectively, demonstrating scalable generation of Rydberg-mediated multiphoton entanglement.

Rydberg interactions can also be exploited for entanglement purification and filtering. Recently, a photon-entanglement filter mediated by Rydberg atoms and capable of operating with noisy input states was experimentally demonstrated\ucite{ye2023photonic}. An initial product state $(|H\rangle_a + |V\rangle_a)(|H\rangle_b + |V\rangle_b)$ consisting of $780~\rm nm$ photons, which contains the target Bell state $|\Psi^+\rangle = |H\rangle_a |V\rangle_b + |V\rangle_a |H\rangle_b$ along with undesired components, was coupled into the atomic ensemble. Horizontal (H) and vertical (V) polarizations were mapped into a spatially separated atomic ensemble through a polarization beam splitter, and coherently stored as collective Rydberg excitations under the $480~\rm nm$ control field, as shown in Fig.~\ref{Fig10}(b). By selecting highly excited Rydberg states with adjacent principal quantum numbers, strong Rydberg interactions can be achieved within the same ensemble. Under the Rydberg blockade mechanism,  double excitations within the same spatial mode (corresponding to $|H\rangle_a |H\rangle_b$ and $|V\rangle_a |V\rangle_b$) are strongly suppressed, whereas the target state $|\Psi^+\rangle$, distributed across the upper and lower ensembles, remains unaffected. After storage, high-fidelity entanglement output of Bell states is achieved through a collective stimulated radiation process. In addition to blockade-based filtering at high principal quantum numbers, dissipative filtering mechanisms were also demonstrated at lower Rydberg levels, where interaction-induced decoherence selectively removes undesired components. Experimentally, entanglement fidelities approaching $99.5\%$ were achieved despite finite storage and retrieval efficiencies, demonstrating the effectiveness of Rydberg-mediated entanglement purification.

Rydberg atomic systems provide a versatile platform for photon–photon entanglement through deterministic emission, sequential multiphoton generation, and entanglement filtering. These approaches extend photonic entanglement from two-photon states toward multiphoton and high-dimensional regimes, offering promising routes toward scalable quantum networks and photonic QIP.

\subsubsection{Atom-atom entanglement}

Deterministic entanglement between neutral atoms is a fundamental requirement for quantum computing and quantum simulation. In neutral atom systems, the realization of strong and controllable interactions has long been a major challenge due to the weak intrinsic coupling between ground-state atoms. The experimental observation of the Rydberg blockade effect provided a practical solution to this limitation, enabling strong and tunable interactions that allow conditional excitation and deterministic entanglement generation. Since its first demonstration, Rydberg-mediated atomic entanglement has evolved from proof-of-principle two-atom experiments toward high-fidelity quantum gates and large-scale programmable atom arrays.

In 2010, Wilk \textit{et al.}\ucite{wilk2010EntanglementTwoIndividual} and Isenhower \textit{et al.}\ucite{isenhower2010demonstration} reported the first experimental demonstrations of two-atom entanglement based on the Rydberg blockade effect. 
The landmark experiment established the practical feasibility of generating deterministic entanglement between individual neutral atoms via Rydberg-mediated interactions, paving the way for fast two-qubit quantum gates. 
To realize scalable quantum computation with Rydberg atoms, achieving high-fidelity entangling gates has become a central objective, prompting the extensive exploration of numerous advanced schemes. For instance, the use of analytical derivative removal by adiabatic gate (DRAG) pulses suppresses energy-level leakage during Rydberg excitation\ucite{theis2016highfidelity}, while global driving with optimized laser detuning and pulse duration improves entanglement fidelity\ucite{han2016implementing}. More generally, advanced pulse optimization and optimal-control methods enhance gate performance under experimental imperfections\ucite{muller2011prospects,goerz2014robustness,saffman2020symmetrica}. In addition, antiblockade and adiabatic protocols improve robustness against parameter fluctuations, ensuring efficient population transfer and entanglement generation\ucite{tian2015populationa,su2017applications,li2018engineering}. Adiabatic schemes, including stimulated Raman adiabatic passage, suppress decoherence and control errors by reducing sensitivity to interatomic distance and temperature\ucite{beterov2016twoqubit,beterov2018adiabatic,idlas2016entanglement,zhao2017robusta,zhang2020submicrosecond}. Interaction engineering via Stark-tuned F\"{o}rster resonances further enables enhanced and controllable interactions for deterministic phase accumulation and high-fidelity multi-qubit gates\ucite{anand2024dualspecies,beterov2018fast}. Beyond coherent gate-based approaches, dissipative engineering exploits controlled dissipation to drive the system into target entangled steady states\ucite{su2015simplified,reiter2016scalable,yang2021dissipative,carr2013preparationa,li2020periodically,rao2014deterministic}, while Rydberg-dressing techniques induce tunable effective interactions with reduced spontaneous emission, enabling long-lived and coherent entanglement\ucite{jau2016Entangling,young2021asymmetric}. These approaches together form a versatile toolbox for robust and scalable Rydberg entanglement generation.

\begin{figure}[htp!]
\centering  
\includegraphics[width=0.99\linewidth]{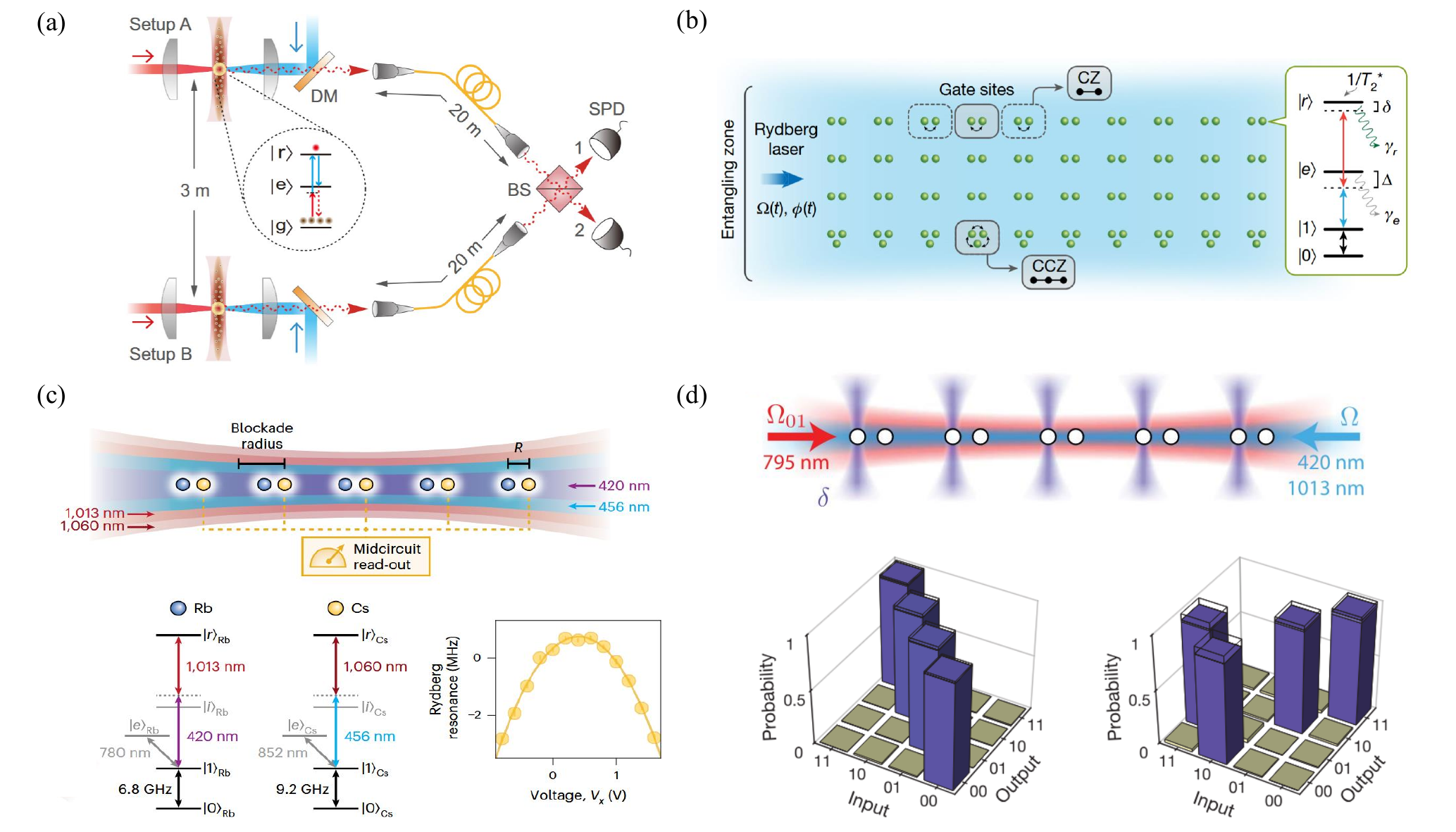}
\caption{(a) Experimental schematic of two independent setups separated by 3~m. Each setup utilizes a microscopic ensemble within a Rydberg blockade sphere to generate remote entanglement (reproduced with permission from Ref.~\cite{yang2025entangling}). (b) Entangling gates are executed by arranging neutral atoms into designated gate sites, where strong interactions are mediated via the Rydberg blockade mechanism driven by global laser pulses (reproduced with permission from Ref.~\cite{evered2023highfidelity}). (c) Arrangement of four focused Rydberg laser beams driving the atom array. Rubidium and cesium atoms are subjected to species-selective two-photon Rydberg excitations (reproduced with permission from Ref.~\cite{anand2024dualspecies}). (d) Atomic pairs globally driven by Raman and Rydberg lasers with local addressing. Following the initialization of four computational basis states, the CNOT gate sequence yields high truth-table fidelities (reproduced with permission from Ref.~\cite{levine2019parallel}).}
\label{Fig11}
\end{figure}

Recent advances have extended Rydberg-mediated entanglement beyond local atomic systems toward long-distance and distributed architectures. In particular, long-distance entanglement between two Rydberg superatoms separated by 3 meters was recently demonstrated\ucite{yang2025entangling}, as shown in Fig.~\ref{Fig11}(a). In this scheme, partial readout of each superatom coherently mapped collective excitations into atom–photon entangled states, which were subsequently interfered to herald remote atomic entanglement. This approach effectively suppressed higher-order excitation noise and achieved robust entanglement with nonzero concurrence across a wide range of readout conditions. In a related development, heralded photon-storage protocols have been employed to establish remote entanglement between spatially separated Rydberg nodes without requiring intermediate interference elements\ucite{an2025entangling}. The measured entanglement fidelity reached $70.5\% \pm 0.6\%$, significantly exceeding the classical threshold. These demonstrations represent important milestones toward scalable quantum networking with neutral atoms, providing practical pathways for implementing distributed quantum repeaters and long-distance quantum communication.

In addition, progress in programmable neutral atom arrays has demonstrated the scalability of Rydberg-mediated entanglement toward large-scale quantum processors. In particular, parallel entanglement operations on Rydberg atom arrays have achieved remarkable success\ucite{evered2023highfidelity}, as shown in Fig.~\ref{Fig11}(b). By optimizing Rydberg laser pulse sequences to realize high-fidelity two-qubit CZ gates, Bell states were generated simultaneously across 20 atoms, achieving an initial Bell-state fidelity of $98.0(2)\%$. Extending this approach to multi-qubit operations, three-qubit CCZ gates were executed in parallel on 21 atoms, enabling the generation of multiple GHZ states with fidelities reaching $90.9(6)\%$. These results highlight the capability of programmable Rydberg arrays to support large-scale, parallel quantum gate operations, representing a significant step toward scalable and fault-tolerant quantum computation. In a related development, coherent transport of entangled atomic pairs within programmable arrays has also been demonstrated\ucite{Bluvstein2024}, with Bell-state fidelities of $94.8(2)\%$ maintained after transport, consistent with the fidelity measured immediately after preparation. Such results demonstrate that entanglement generation and coherent manipulation can be reliably preserved across extended atomic systems, further reinforcing the scalability of neutral atom architectures.

Beyond single-species platforms, diversified atomic architectures have also shown strong potential for scalable quantum technologies. Dual-species atomic arrays, for example, provide additional flexibility for interaction control and state manipulation. An earlier step toward heterogeneous neutral-atom architectures was demonstrated by Zeng \textit{et al.}, who realized deterministic Bell-state generation between individually trapped $^{87}$Rb and $^{85}$Rb atoms, achieving raw fidelities of $0.73(1)$ and $0.59(3)$, respectively\ucite{zeng2017entangling}. In a rubidium–cesium dual-species array, interspecies entanglement with a fidelity of $0.69(3)$ was achieved using tunable Rydberg–F{\"o}rster resonances\ucite{anand2024dualspecies}, as shown in Fig.~\ref{Fig11}(c). Rydberg arrays have also demonstrated reliable generation of multi-atom GHZ states, with fidelities exceeding $50\%$ after SPAM correction and two-qubit Bell-state fidelities reaching $95.5\%$\ucite{graham2022multiqubit}.  In addition, alkaline-earth atomic systems with more complex electronic structures offer new opportunities for high-coherence entanglement. For instance, an entanglement fidelity of $F=0.991(4)$ has been reported for a pair of strontium ($^{88}$Sr) atoms\ucite{madjarov2020highfidelity}, highlighting the potential of these platforms for precision quantum control. Furthermore, global-control protocols based on tailored laser detuning and phase modulation have enabled parallel entanglement of multiple atomic pairs without requiring individual addressing\ucite{levine2019parallel,levine2018highfidelity}, achieving calibrated fidelities up to $97.4(3)\%$, as shown in Fig.~\ref{Fig11}(d). These developments collectively demonstrate the versatility of Rydberg-based systems across different atomic species and control strategies. The representative experimental summary of Rydberg-mediated atom-atom entanglement is shown in Table \ref{Table3}.

\begin{table}[htp!]
\centering
\small
\renewcommand{\arraystretch}{1.0}
\setlength{\tabcolsep}{12pt}
\begin{tabular}{@{}c l  c c c@{}}
\hline
Year & Type  & Fidelity &
\shortstack{determinism} & Reference \\
\hline
2010 & Bell states  
& 0.75 
& 0.61 
&\textit{Phys. Rev. Lett.} \ucite{wilk2010EntanglementTwoIndividual} \\

2010 & CNOT 
& 0.73 
& 1 
& \textit{Phys. Rev. Lett.} \ucite{isenhower2010demonstration} \\

2016 & Bell states 
& 0.81(2) 
& 0.60(3) 
& \textit{Nat. Phys.} \ucite{jau2016Entangling} \\

2017 & CNOT/Bell states 
& 0.73(1)/0.59(3) 
& 1 
& \textit{Phys. Rev. Lett.} \ucite{zeng2017entangling} \\

2018 & Bell states 
& 0.97(3) 
& 1 
& \textit{Phys. Rev. Lett.} \ucite{levine2018highfidelity} \\

2019 & CZ/Toffoli 
& 0.974(3) 
& 1 
& \textit{Phys. Rev. Lett.} \ucite{levine2019parallel} \\

2020 & Bell states 
& 0.78 
& -- 
& \textit{Nature} \ucite{zhang2020submicrosecond} \\

2020 & Sr Bell states 
& 0.991(4) 
& 1 
& \textit{Nat. Phys.} \ucite{madjarov2020highfidelity} \\

2022 & Bell/GHZ states 
& 0.955 
& 1 
& \textit{Nature} \ucite{graham2022multiqubit} \\

2023 & CZ/CCZ 
& 0.9952(4) 
& 1 
& \textit{Nature} \ucite{evered2023highfidelity} \\

2024 & Rb--Cs Bell states 
& 0.69(3) 
& 1 
& \textit{Nat. Phys.} \ucite{anand2024dualspecies} \\

2024 & Transported Bell states  
& 0.948(2) 
& 1 
& \textit{Nature} \ucite{Bluvstein2024} \\

2025 & Remote Bell states 
& -- 
& -- 
& \textit{Phys. Rev. Lett.} \ucite{yang2025entangling} \\

2025 & Heralded Bell states 
& 0.705(6) 
& -- 
& \textit{Phys. Rev. Lett.} \ucite{an2025entangling} \\
\hline
\end{tabular}
\caption{Summary of Rydberg-mediated atom--atom entanglement experiments.}
\label{Table3}
\end{table}

Due to the complexity of manipulating high-lying Rydberg excitations and the required coherent retrieval from polaritons back to photons, usually only one type of Rydberg state is used for storing photons. However, if multiple Rydberg states are employed for storing single photons, it is possible to engineer highly correlated polariton pairs with specific strengths offered by appropriate superpositions of different Rydberg states\ucite{PhysRevA.95.043429,PhysRevA.97.033414}, where an analogue of spin-charge separation can emerge as predicted in Ref.~\cite{ShiJPB2016}.

Rydberg-mediated interactions provide a powerful and scalable platform for deterministic atomic entanglement, ranging from early two-atom demonstrations to high-fidelity quantum gates and large-scale programmable arrays. Recent advances in distributed entanglement and multi-species platforms further extend the applicability of Rydberg systems toward scalable quantum networks and fault-tolerant quantum computing architectures.

\section{Conclusion and Outlook}\label{sec:4}

In this review, we have presented a systematic overview of Rydberg-mediated nonlinear quantum optics. We have outlined the fundamental principles underlying Rydberg-mediated nonlinear quantum optics, including the unique scaling properties of Rydberg atoms, the mechanism of Rydberg blockade, and the formation of collective excitations. By employing a Rydberg-EIT scheme, strong Rydberg–Rydberg interactions are efficiently mapped onto photons via the formation of Rydberg polaritons, enabling effective photon–photon interactions at the single-photon level. This capability overcomes the intrinsic weakness of conventional optical nonlinearities and establishes Rydberg systems as a powerful platform for quantum optics. Such a framework has enabled a broad range of advances, including deterministic single-photon sources and coherent manipulation schemes, single-photon switches and transistors, photonic quantum gates, and a rich variety for entanglement generation. Notably, the emergence of contactless nonlinear optics and nonlocal photon-photon interactions extend the paradigm beyond traditional spatial constraints, bringing new perspectives for distributed and modular quantum architectures.

Despite the remarkable achievements of Rydberg-mediated nonlinear quantum optics, its performance is ultimately constrained by several competing physical mechanisms, including limited optical depth per blockade volume ($\mathrm{OD}_b$), low photon storage efficiency, limited scalability, rapid many-body dephasing with multiple-photon polaritons for contactless photon-photon interaction, and motion-induced dephasing. Efficient photon storage and strong photon-photon interactions both require a large $\mathrm{OD}_b$, which increases the collective light-matter coupling and enhances nonlinear effects. However, increasing the atomic density to improve $\mathrm{OD}_b$ also leads to stronger atomic collisions, radiation trapping, and interaction-induced dephasing, thereby reducing the storage lifetime and retrieval efficiency. Likewise, stronger Rydberg interactions, obtained by exciting higher principal quantum numbers, enlarge the blockade radius and strengthen optical nonlinearities, but simultaneously increase the sensitivity to electric-field noise, blackbody radiation, and spontaneous decay. Furthermore, a stronger control field broadens the EIT transparency window and facilitates efficient pulse propagation, whereas a weaker control field is preferred for achieving slower group velocities and higher photonic compression. These competing requirements define the practical operating regime of Rydberg-mediated nonlinear quantum optics and motivate ongoing efforts toward optimized atomic trapping, cavity enhancement, and hybrid quantum architectures.

To address these limitations, several key strategies are being actively developed, including extending motional coherence times through magic-wavelength optical trapping\ucite{lampen2018Longliveda,zhang2011magicwavelength,wilson2022trapping}, state-mapping techniques\ucite{PhysRevLett.134.053604,li2026,shi2025coherence,shi2020suppressing}, exploring potential strengths of alternative atomic species\ucite{bbv3-d4ch}, enhancing light–matter coupling through integration with high-finesse optical cavities\ucite{sheng2017intracavity,guerlin2010cavity}, and improving scalability by implementing high-efficiency atomic loading\ucite{manetsch2025tweezer,pichard2024rearrangement} and low-crosstalk addressing methods\ucite{radnaev2025universal,kaufman2021quantum}. With continued progress along these directions, the scope of Rydberg-mediated nonlinear quantum optics will transcend the few-photon limit, future explorations into multi-photon quantum optics will enable the synthesis of exotic many-body states of light—such as photonic Wigner crystals, fractional quantum Hall states of light, and drive the development of robust multi-photon quantum devices. Furthermore, the ability to efficiently interface flying photons with stationary Rydberg excitations will facilitate the implementation of remote quantum logic operations between distant modules, providing a crucial hardware foundation for large-scale distributed quantum computing and quantum internet.

Moreover, hybridizing Rydberg systems with other state-of-the-art quantum platforms will unlock transformative capabilities. Integrating these systems with reconfigurable single-atom tweezer arrays will facilitate scalable quantum computation and error correction, while their coupling with superconducting circuits offers a highly efficient route for coherent, single-photon-level quantum transduction between microwave and optical frequency bands. 
Owing to the giant electric dipole moments of Rydberg states and the coherent optical interface provided by Rydberg EIT, microwave photons can be coherently converted into optical photons through nonlinear wave-mixing processes. Such Rydberg-based quantum transducers provide a promising interface between superconducting quantum processors and optical quantum networks, enabling long-distance distribution of quantum information while preserving the advantages of optical communication. Significant theoretical and experimental progress has been achieved in recent years, including coherent microwave-to-optical conversion via six-wave mixing, high-efficiency transduction, and hybrid superconducting-atom architectures\ucite{han2018coherent,tu2022highefficiency,borowka2024continuous,petrosyan2019microwave}. Although a comprehensive discussion is beyond the scope of the present review, we believe this rapidly developing direction will play an increasingly important role in future hybrid quantum networks.

\addcontentsline{toc}{chapter}{Acknowledgment}
\section*{Acknowledgment}
This work is supported by the National Natural Science Foundation of China  (No. 12241408, U2341211, 12120101004, 12504304, and 12547103); Changjiang Scholars and Innovative Research  Team in University of Ministry of Education of China (No. IRT\_17R70); and Shanxi Province's Basic  Research Program (202503021212074).

\renewcommand{\refname}{References}
\bibliographystyle{iopart-num.bst}

\begin{thebibliography}{100}
\expandafter\ifx\csname url\endcsname\relax
  \def\url#1{{\tt #1}}\fi
\expandafter\ifx\csname urlprefix\endcsname\relax\def\urlprefix{URL }\fi
\providecommand{\eprint}[2][]{\url{#2}}

\bibitem{mabuchi2002cavity}
Mabuchi H and Doherty A~C 2002 {\em Science\/} {\bf 298} 1372--1377

\bibitem{birnbaum2005photon}
Birnbaum K~M, Boca A, Miller R, Boozer A~D, Northup T~E and Kimble H~J 2005 {\em Nature\/} {\bf 436} 87--90

\bibitem{dayan2008photon}
Dayan B, Parkins A~S, Aoki T, Ostby E~P, Vahala K~J and Kimble H~J 2008 {\em Science\/} {\bf 319} 1062--1065

\bibitem{kimble2008quantum}
Kimble H~J 2008 {\em Nature\/} {\bf 453} 1023--1030

\bibitem{chen2013alloptical}
Chen W, Beck K~M, B{\"u}cker R, Gullans M, Lukin M~D, {Tanji-Suzuki} H and Vuleti{\'c} V 2013 {\em Science\/} {\bf 341} 768--770

\bibitem{reiserer2015cavitybased}
Reiserer A and Rempe G 2015 {\em Rev. Mod. Phys.\/} {\bf 87} 1379--1418

\bibitem{zwanenburg2013silicon}
Zwanenburg F~A, Dzurak A~S, Morello A, Simmons M~Y, Hollenberg L~C~L, Klimeck G, Rogge S, Coppersmith S~N and Eriksson M~A 2013 {\em Rev. Mod. Phys.\/} {\bf 85}(3) 961--1019

\bibitem{lu2021quantumdot}
Lu C~Y and Pan J~W 2021 {\em Nat. Nanotechnol.\/} {\bf 16} 1294--1296

\bibitem{lodahl2015interfacing}
Lodahl P, Mahmoodian S and Stobbe S 2015 {\em Rev. Mod. Phys.\/} {\bf 87}(2) 347--400

\bibitem{senellart2017highperformance}
Senellart P, Solomon G and White A 2017 {\em Nat. Nanotechnol.\/} {\bf 12} 1026--1039

\bibitem{blatt2008entangled}
Blatt R and Wineland D 2008 {\em Nature\/} {\bf 453} 1008--1015

\bibitem{monroe2013scaling}
Monroe C and Kim J 2013 {\em Science\/} {\bf 339} 1164--1169

\bibitem{bruzewicz2019trappedion}
Bruzewicz C~D, Chiaverini J, McConnell R and Sage J~M 2019 {\em Applied Physics Reviews\/} {\bf 6} 021314

\bibitem{doherty2013nitrogenvacancy}
Doherty M~W, Manson N~B, Delaney P, Jelezko F, Wrachtrup J and Hollenberg L~C 2013 {\em Physics Reports\/} {\bf 528} 1--45

\bibitem{zhou2014quantum}
Zhou J~W, Wang P~F, Shi F~Z, Huang P, Kong X, Xu X~K, Zhang Q, Wang Z~X, Rong X and Du J~F 2014 {\em Front. Phys.\/} {\bf 9} 587--597

\bibitem{awschalom2018quantum}
Awschalom D~D, Hanson R, Wrachtrup J and Zhou B~B 2018 {\em Nature Photonics\/} {\bf 12} 516--527

\bibitem{song2019generation}
Song C, Xu K, Li H, Zhang Y~R, Zhang X, Liu W, Guo Q, Wang Z, Ren W, Hao J, Feng H, Fan H, Zheng D, Wang D~W, Wang H and Zhu S~Y 2019 {\em Science\/} {\bf 365} 574--577

\bibitem{ruf2021quantum}
Ruf M, Wan N~H, Choi H, Englund D and Hanson R 2021 {\em Journal of Applied Physics\/} {\bf 130} 070901

\bibitem{devoret2013superconducting}
Devoret M~H and Schoelkopf R~J 2013 {\em Science\/} {\bf 339} 1169--1174

\bibitem{blais2021circuit}
Blais A, Grimsmo A~L, Girvin S~M and Wallraff A 2021 {\em Reviews of Modern Physics\/} {\bf 93} 025005

\bibitem{gu2017microwave}
Gu X, Kockum A~F, Miranowicz A, Liu Y~x and Nori F 2017 {\em Physics Reports\/} {\bf 718--719} 1--102

\bibitem{saffman2010quantum}
Saffman M, Walker T~G and M{\o}lmer K 2010 {\em Rev. Mod. Phys.\/} {\bf 82} 2313--2363

\bibitem{firstenberg2016nonlinear}
Firstenberg O, Adams C~S and Hofferberth S 2016 {\em J. Phys. B: At. Mol. Opt. Phys.\/} {\bf 49} 152003

\bibitem{adams2020rydberg}
Adams C~S, Pritchard J~D and Shaffer J~P 2020 {\em J. Phys. B: At. Mol. Opt. Phys.\/} {\bf 53} 012002

\bibitem{gallagher1994rydberg}
Gallagher T~F 1994 {\em {Rydberg} {{Atoms}}\/} 1st ed Cambridge {{Monographs}} on {{Atomic}}, {{Molecular}} and {{Chemical Physics}} (Cambridge: Cambridge University Press)

\bibitem{sibalic2018rydberg}
{\v S}ibali{\'c} N and Adams C~S 2018 {\em {Rydberg} {{Physics}}\/} (IOP Publishing)

\bibitem{shao2024rydberg}
Shao X~Q, Su S~L, Li L, Nath R, Wu J~H and Li W 2024 {\em Applied Physics Reviews\/} {\bf 11} 031320

\bibitem{fleischhauer2000darkstate}
Fleischhauer M and Lukin M~D 2000 {\em Phys. Rev. Lett.\/} {\bf 84} 5094--5097

\bibitem{fleischhauer2002quantum}
Fleischhauer M and Lukin M~D 2002 {\em Phys. Rev. A\/} {\bf 65} 022314

\bibitem{fleischhauer2005electromagnetically}
Fleischhauer M, Imamoglu A and Marangos J~P 2005 {\em Rev. Mod. Phys.\/} {\bf 77} 633--673

\bibitem{friedler2005longrange}
Friedler I, Petrosyan D, Fleischhauer M and Kurizki G 2005 {\em Phys. Rev. A\/} {\bf 72} 043803

\bibitem{gorshkov2011photonphoton}
Gorshkov A~V, Otterbach J, Fleischhauer M, Pohl T and Lukin M~D 2011 {\em Phys. Rev. Lett.\/} {\bf 107} 133602

\bibitem{chang2014quantum}
Chang D~E, Vuleti{\'c} V and Lukin M~D 2014 {\em Nat. Photon.\/} {\bf 8} 685--694

\bibitem{boyd2020nonlinear}
Boyd R~W 2020 {\em Nonlinear {{Optics}}\/} 4th ed (San Diego: Elsevier Science \& Technology)

\bibitem{lukin2001dipole}
Lukin M~D, Fleischhauer M, Cote R, Duan L~M, Jaksch D, Cirac J~I and Zoller P 2001 {\em Phys. Rev. Lett.\/} {\bf 87} 037901

\bibitem{tong2004local}
Tong D, Farooqi S~M, Stanojevic J, Krishnan S, Zhang Y~P, C{\^o}t{\'e} R, Eyler E~E and Gould P~L 2004 {\em Phys. Rev. Lett.\/} {\bf 93} 063001

\bibitem{gaetan2009observation}
Ga{\"e}tan A, Miroshnychenko Y, Wilk T, Chotia A, Viteau M, Comparat D, Pillet P, Browaeys A and Grangier P 2009 {\em Nat. Phys.\/} {\bf 5} 115--118

\bibitem{pritchard2010cooperative}
Pritchard J~D, Maxwell D, Gauguet A, Weatherill K~J, Jones M~P~A and Adams C~S 2010 {\em Phys. Rev. Lett.\/} {\bf 105} 193603

\bibitem{kazemi2023drivendissipative}
Kazemi J and Weimer H 2023 {\em Phys. Rev. Lett.\/} {\bf 130} 163601

\bibitem{vuletic2006when}
Vuletic V 2006 {\em Nat. Phys.\/} {\bf 2} 801--802

\bibitem{heidemann2007evidence}
Heidemann R, Raitzsch U, Bendkowsky V, Butscher B, L{\"o}w R, Santos L and Pfau T 2007 {\em Phys. Rev. Lett.\/} {\bf 99} 163601

\bibitem{paris-mandoki2017freespace}
{Paris-Mandoki} A, Braun C, Kumlin J, Tresp C, Mirgorodskiy I, Christaller F, B{\"u}chler H~P and Hofferberth S 2017 {\em Phys. Rev. X\/} {\bf 7} 041010

\bibitem{kumlin2023quantum}
{kumlin} J, Braun C, Tresp C, Stiesdal N, Hofferberth S and {Paris-Mandoki} A 2023 {\em J. Phys. Commun.\/} {\bf 7} 052001

\bibitem{dudin2012observation}
Dudin Y~O, Li L, Bariani F and Kuzmich A 2012 {\em Nat. Phys.\/} {\bf 8} 790--794

\bibitem{wilk2010entanglement}
Wilk T, Ga{\"e}tan A, Evellin C, Wolters J, Miroshnychenko Y, Grangier P and Browaeys A 2010 {\em Phys. Rev. Lett.\/} {\bf 104} 010502

\bibitem{saffman2016quantum}
Saffman M 2016 {\em J. Phys. B: At. Mol. Opt. Phys.\/} {\bf 49} 202001

\bibitem{zeng2017entangling}
Zeng Y, Xu P, He X, Liu Y, Liu M, Wang J, Papoular D~J, Shlyapnikov G~V and Zhan M 2017 {\em Physical Review Letters\/} {\bf 119} 160502

\bibitem{madjarov2020highfidelity}
Madjarov I~S, Covey J~P, Shaw A~L, Choi J, Kale A, Cooper A, Pichler H, Schkolnik V, Williams J~R and Endres M 2020 {\em Nat. Phys.\/} {\bf 16} 857--861

\bibitem{levine2018highfidelity}
Levine H, Keesling A, Omran A, Bernien H, Schwartz S, Zibrov A~S, Endres M, Greiner M, Vuleti{\'c} V and Lukin M~D 2018 {\em Phys. Rev. Lett.\/} {\bf 121} 123603

\bibitem{levine2019parallel}
Levine H, Keesling A, Semeghini G, Omran A, Wang T~T, Ebadi S, Bernien H, Greiner M, Vuleti{\'c} V, Pichler H and Lukin M~D 2019 {\em Phys. Rev. Lett.\/} {\bf 123} 170503

\bibitem{anand2024dualspecies}
Anand S, Bradley C~E, White R, Ramesh V, Singh K and Bernien H 2024 {\em Nat. Phys.\/} {\bf 20} 1744--1750

\bibitem{graham2022multiqubit}
Graham T~M, Song Y, Scott J and \emph{et al} 2022 {\em Nature\/} {\bf 604} 457--462

\bibitem{evered2023highfidelity}
Evered S~J, Bluvstein D, Kalinowski M, Ebadi S, Manovitz T, Zhou H, Li S~H, Geim A~A, Wang T~T, Maskara N, Levine H, Semeghini G, Greiner M, Vuleti{\'c} V and Lukin M~D 2023 {\em Nature\/} {\bf 622} 268--272

\bibitem{Bluvstein2024}
Bluvstein D, Evered S~J, Geim A~A and \emph{et al} 2024 {\em Nature\/} {\bf 626} 58--65

\bibitem{dudin2012Strongly}
Dudin Y~O and Kuzmich A 2012 {\em Science\/} {\bf 336} 887--889

\bibitem{peyronel2012quantum}
Peyronel T, Firstenberg O, Liang Q~Y, Hofferberth S, Gorshkov A~V, Pohl T, Lukin M~D and Vuleti{\'c} V 2012 {\em Nature\/} {\bf 488} 57--60

\bibitem{li2016Quantuma}
Li L and Kuzmich A 2016 {\em Nat. Commun.\/} {\bf 7} 13618

\bibitem{petrosyan2018deterministic}
Petrosyan D and M{\o}lmer K 2018 {\em Phys. Rev. Lett.\/} {\bf 121} 123605

\bibitem{ornelas-huerta2020Ondemand}
{Ornelas-Huerta} D~P, Craddock A~N, Goldschmidt E~A, Hachtel A~J, Wang Y, Bienias P, Gorshkov A~V, Rolston S~L and Porto J~V 2020 {\em Optica\/} {\bf 7} 813

\bibitem{li2026}
Li C, Shi X~F, Jiao Y, Shen X, Yang J, Bai J, Adams C, Jia S and Zhao J 2026 {\em Optica\/} {\bf 13} 914--919

\bibitem{baur2014SinglePhotonb}
Baur S, Tiarks D, Rempe G and D{\"u}rr S 2014 {\em Phys. Rev. Lett.\/} {\bf 112} 073901

\bibitem{gorniaczyk2014SinglePhoton}
Gorniaczyk H, Tresp C, Schmidt J, Fedder H and Hofferberth S 2014 {\em Phys. Rev. Lett.\/} {\bf 113} 053601

\bibitem{tiarks2014SinglePhoton}
Tiarks D, Baur S, Schneider K, D{\"u}rr S and Rempe G 2014 {\em Phys. Rev. Lett.\/} {\bf 113} 053602

\bibitem{gorniaczyk2016enhancement}
Gorniaczyk H, Tresp C, Bienias P, {Paris-Mandoki} A, Li W, Mirgorodskiy I, B{\"u}chler H~P, Lesanovsky I and Hofferberth S 2016 {\em Nat. Commun.\/} {\bf 7} 12480

\bibitem{liao2025nonlocal}
Liao R, Song Z~R, Ye G~S, Yu J~H, Chang Y and Li L 2025 {\em Phys. Rev. Lett.\/} {\bf 135} 260803

\bibitem{tiarks2016optical}
Tiarks D, Schmidt S, Rempe G and D{\"u}rr S 2016 {\em Sci. Adv.\/} {\bf 2} e1600036

\bibitem{tiarks2019Photon}
Tiarks D, {Schmidt-Eberle} S, Stolz T, Rempe G and D{\"u}rr S 2019 {\em Nat. Phys.\/} {\bf 15} 124--126

\bibitem{stolz2022quantumlogic}
Stolz T, Hegels H, Winter M, R{\"o}hr B, Hsiao Y~F, Husel L, Rempe G and D{\"u}rr S 2022 {\em Phys. Rev. X\/} {\bf 12} 021035

\bibitem{an2025entangling}
An Z~Y, Lu B~W, Li J, Yang C~W, Li L, Bao X~H and Pan J~W 2025 {\em Phys. Rev. Lett.\/} {\bf 134} 230803

\bibitem{firstenberg2013attractive}
Firstenberg O, Peyronel T, Liang Q~Y, Gorshkov A~V, Lukin M~D and Vuleti{\'c} V 2013 {\em Nature\/} {\bf 502} 71--75

\bibitem{liang2018observation}
Liang Q~Y, Venkatramani A~V, Cantu S~H, Nicholson T~L, Gullans M~J, Gorshkov A~V, Thompson J~D, Chin C, Lukin M~D and Vuleti{\'c} V 2018 {\em Science\/} {\bf 359} 783--786

\bibitem{otterbach2013wigner}
Otterbach J, Moos M, Muth D and Fleischhauer M 2013 {\em Phys. Rev. Lett.\/} {\bf 111} 113001

\bibitem{busche2017Contactlessb}
Busche H, Huillery P, Ball S~W, Ilieva T, Jones M~P~A and Adams C~S 2017 {\em Nat. Phys.\/} {\bf 13} 655--658

\bibitem{robertson2021arc}
Robertson E, {\v S}ibali{\'c} N, Potvliege R and Jones M 2021 {\em Comput. Phys. Commun.\/} {\bf 261} 107814

\bibitem{low2012experimental}
L{\"o}w R, Weimer H, Nipper J, Balewski J~B, Butscher B, B{\"u}chler H~P and Pfau T 2012 {\em J. Phys. B: At. Mol. Opt. Phys.\/} {\bf 45} 113001

\bibitem{PhysRevA.77.032723}
Walker T~G and Saffman M 2008 {\em Phys. Rev. A\/} {\bf 77}(3) 032723

\bibitem{shi_quantum_2022}
Shi X~F 2022 {\em Quantum Sci. Technol.\/} {\bf 7} 023002

\bibitem{barredo2015coherent}
Barredo D, Labuhn H, Ravets S, Lahaye T, Browaeys A and Adams C~S 2015 {\em Phys. Rev. Lett.\/} {\bf 114} 113002

\bibitem{labuhn2016tunable}
Labuhn H, Barredo D, Ravets S, De~L{\'e}s{\'e}leuc S, Macr{\`i} T, Lahaye T and Browaeys A 2016 {\em Nature\/} {\bf 534} 667--670

\bibitem{khazali2019polariton}
Khazali M, Murray C~R and Pohl T 2019 {\em Phys. Rev. Lett.\/} {\bf 123} 113605

\bibitem{urban2009observation}
Urban E, Johnson T~A, Henage T, Isenhower L, Yavuz D~D, Walker T~G and Saffman M 2009 {\em Nat. Phys.\/} {\bf 5} 110--114

\bibitem{honer2010collective}
Honer J, Weimer H, Pfau T and B{\"u}chler H~P 2010 {\em Phys. Rev. Lett.\/} {\bf 105} 160404

\bibitem{pohl2010dynamical}
Pohl T, Demler E and Lukin M~D 2010 {\em Phys. Rev. Lett.\/} {\bf 104} 043002

\bibitem{weimer2010rydberg}
Weimer H, M{\"u}ller M, Lesanovsky I, Zoller P and B{\"u}chler H~P 2010 {\em Nat. Phys.\/} {\bf 6} 382--388

\bibitem{wu2021concise}
Wu X, Liang X, Tian Y, Yang F, Chen C, Liu Y~C, Tey M~K and You L 2021 {\em Chinese Phys. B\/} {\bf 30} 020305

\bibitem{jiao2020singlephoton}
Jiao Y, Spong N~L~R, Hughes O~D~W, So C, Ilieva T, Weatherill K~J and Adams C~S 2020 {\em Opt. Lett.\/} {\bf 45} 5888

\bibitem{honer2011artificial}
Honer J, L{\"o}w R, Weimer H, Pfau T and B{\"u}chler H~P 2011 {\em Phys. Rev. Lett.\/} {\bf 107} 093601

\bibitem{mohapatra2007coherent}
Mohapatra A~K, Jackson T~R and Adams C~S 2007 {\em Phys. Rev. Lett.\/} {\bf 98} 113003

\bibitem{weatherill2008electromagnetically}
Weatherill K~J, Pritchard J~D, Abel R~P, Bason M~G, Mohapatra A~K and Adams C~S 2008 {\em J. Phys. B: At. Mol. Opt. Phys.\/} {\bf 41} 201002

\bibitem{petrosyan2011electromagnetically}
Petrosyan D, Otterbach J and Fleischhauer M 2011 {\em Phys. Rev. Lett.\/} {\bf 107} 213601

\bibitem{budker1999nonlinear}
Budker D, Kimball D~F, Rochester S~M and Yashchuk V~V 1999 {\em Phys. Rev. Lett.\/} {\bf 83} 1767--1770

\bibitem{novikova2012electromagnetically}
Novikova I, Walsworth R and Xiao Y 2012 {\em Laser \& Photonics Reviews\/} {\bf 6} 333--353

\bibitem{kash1999ultraslow}
Kash M~M, Sautenkov V~A, Zibrov A~S, Hollberg L, Welch G~R, Lukin M~D, Rostovtsev Y, Fry E~S and Scully M~O 1999 {\em Phys. Rev. Lett.\/} {\bf 82} 5229--5232

\bibitem{hau1999light}
Hau L~V, Harris S~E, Dutton Z and Behroozi C~H 1999 {\em Nature\/} {\bf 397} 594--598

\bibitem{distante2017storing}
Distante E, Farrera P, Padr{\'o}n-Brito A, Paredes-Barato D, Heinze G and De~Riedmatten H 2017 {\em Nat. Commun.\/} {\bf 8} 14072

\bibitem{jiao2025suppression}
Jiao Y, Li C, Shi X~F, Fan J, Bai J, Jia S, Zhao J and Adams C~S 2025 {\em Phys. Rev. Lett.\/} {\bf 134} 053604

\bibitem{distante2016storage}
Distante E, {Padr{\'o}n-Brito} A, Cristiani M, {Paredes-Barato} D and De~Riedmatten H 2016 {\em Phys. Rev. Lett.\/} {\bf 117} 113001

\bibitem{schmidt-eberle2020darktime}
{Schmidt-Eberle} S, Stolz T, Rempe G and D{\"u}rr S 2020 {\em Phys. Rev. A\/} {\bf 101} 013421

\bibitem{maring2024versatile}
Maring N, Fyrillas A, Pont M and \emph{et al} 2024 {\em Nat. Photon.\/} {\bf 18} 603--609

\bibitem{obrien2007optical}
O'Brien J~L 2007 {\em Science\/} {\bf 318} 1567--1570

\bibitem{couteau2023applications}
Couteau C, Barz S, Durt T, Gerrits T, Huwer J, Prevedel R, Rarity J, Shields A and Weihs G 2023 {\em Nat. Rev. Phys.\/} {\bf 5} 326--338

\bibitem{aspuru-guzik2012photonic}
{Aspuru-Guzik} A and Walther P 2012 {\em Nat. Phys.\/} {\bf 8} 285--291

\bibitem{hartmann2016quantum}
Hartmann M~J 2016 {\em J. Opt.\/} {\bf 18} 104005

\bibitem{hu2016experimental}
Hu J~Y, Yu B, Jing M~Y, Xiao L~T, Jia S~T, Qin G~Q and Long G~L 2016 {\em Light Sci. Appl.\/} {\bf 5} e16144--e16144

\bibitem{li2020quantum}
Li T and Long G~L 2020 {\em New J. Phys.\/} {\bf 22} 063017

\bibitem{qi2019implementation}
Qi R, Sun Z, Lin Z, Niu P, Hao W, Song L, Huang Q, Gao J, Yin L and Long G~L 2019 {\em Light Sci. Appl.\/} {\bf 8} 22

\bibitem{pan2012multiphoton}
Pan J~W, Chen Z~B, Lu C~Y, Weinfurter H, Zeilinger A and \ifmmode~\dot{Z}\else \.{Z}\fi{}ukowski M 2012 {\em Rev. Mod. Phys.\/} {\bf 84}(2) 777--838

\bibitem{guo2023ultrathin}
Guo Q, Qi X~Z, Zhang L and \emph{et al} 2023 {\em Nature\/} {\bf 613} 53--59

\bibitem{kaneda2016heralded}
Kaneda F, {Garay-Palmett} K, U'Ren A~B and Kwiat P~G 2016 {\em Opt. Express\/} {\bf 24} 10733

\bibitem{gallagher2008dipole}
Gallagher T~F and Pillet P 2008 Dipole--{{Dipole Interactions}} of {{Rydberg Atoms}} {\em Advances {{In Atomic}}, {{Molecular}}, and {{Optical Physics}}\/} ({\em Advances in {{Atomic}}, {{Molecular}}, and {{Optical Physics}}\/} vol~56) (Academic Press) pp 161--218

\bibitem{browaeys2016experimental}
Browaeys A, Barredo D and Lahaye T 2016 {\em J. Phys. B: At. Mol. Opt. Phys.\/}

\bibitem{giudici2025fasta}
Giudici G, Veroni S, Giudice G, Pichler H and Zeiher J 2025 {\em PRX Quantum\/} {\bf 6} 030308

\bibitem{deleseleuc2017optical}
De~L{\'e}s{\'e}leuc S, Barredo D, Lienhard V, Browaeys A and Lahaye T 2017 {\em Phys. Rev. Lett.\/} {\bf 119} 053202

\bibitem{PhysRevA.85.033811}
Bariani F and Kennedy T~A~B 2012 {\em Phys. Rev. A\/} {\bf 85}(3) 033811

\bibitem{PhysRevA.86.041802}
Bariani F, Goldbart P~M and Kennedy T~A~B 2012 {\em Phys. Rev. A\/} {\bf 86}(4) 041802(R)

\bibitem{Bariani2012}
Bariani F, Dudin Y~O, Kennedy T~A~B and Kuzmich A 2012 {\em Phys. Rev. Lett.\/} {\bf 108} 030501

\bibitem{Maxwell2013}
Maxwell D, Szwer D~J, Paredes-Barato D, Busche H, Pritchard J~D, Gauguet A, Weatherill K~J, Jones M~P~A and Adams C~S 2013 {\em Phys. Rev. Lett.\/} {\bf 110}(10) 103001

\bibitem{maxwell2014microwave}
Maxwell D, Szwer D~J, {Paredes-Barato} D, Busche H, Pritchard J~D, Gauguet A, Jones M~P~A and Adams C~S 2014 {\em Phys. Rev. A\/} {\bf 89} 043827

\bibitem{spong2021collectively}
Spong N~L~R, Jiao Y, Hughes O~D~W, Weatherill K~J, Lesanovsky I and Adams C~S 2021 {\em Phys. Rev. Lett.\/} {\bf 127} 063604

\bibitem{magro2023deterministic}
Magro V, Vaneecloo J, Garcia S and Ourjoumtsev A 2023 {\em Nat. Photon.\/} {\bf 17} 688--693

\bibitem{xu2024continuously}
Xu B, Ye G~S, Chang Y, Shi T and Li L 2024 {\em Rep. Prog. Phys.\/} {\bf 87} 110502

\bibitem{fan2023manipulation}
Fan J, Zhang H, Jiao Y, Li C, Bai J, Wu J, Zhao J and Jia S 2023 {\em Opt. Express\/} {\bf 31} 20641

\bibitem{fan2023robust}
Fan J, Jiao Y, Li C, Bai J, Zhao J and Jia S 2023 {\em Opt. Express\/} {\bf 31} 31654

\bibitem{xu2021fast}
Xu W, Venkatramani A~V, Cant{\'u} S~H, {\v S}umarac T, Kl{\"u}sener V, Lukin M~D and Vuleti{\'c} V 2021 {\em Phys. Rev. Lett.\/} {\bf 127} 050501

\bibitem{chen2021twocolor}
Chen C, Yang F, Wu X, Shen C, Tey M~K and You L 2021 {\em Phys. Rev. A\/} {\bf 103} 053303

\bibitem{hao2019singlephoton}
Hao Y~M, Lin G~W, Lin X~M, Niu Y~P and Gong S~Q 2019 {\em Sci. Rep.\/} {\bf 9} 4723

\bibitem{ding2023facilitationinduced}
Ding Y, Bai Z, Huang G and Li W 2023 {\em Phys. Rev. Appl.\/} {\bf 19} 014017

\bibitem{wehner2018quantum}
Wehner S, Elkouss D and Hanson R 2018 {\em Science\/} {\bf 362} eaam9288

\bibitem{azuma2023quantum}
Azuma K, Economou S~E, Elkouss D, Hilaire P, Jiang L, Lo H~K and Tzitrin I 2023 {\em Rev. Mod. Phys.\/} {\bf 95} 045006

\bibitem{lloyd1995almost}
Lloyd S 1995 {\em Phys. Rev. Lett.\/} {\bf 75} 346--349

\bibitem{sleator1995realizable}
Sleator T and Weinfurter H 1995 {\em Phys. Rev. Lett.\/} {\bf 74} 4087--4090

\bibitem{monroe1995demonstration}
Monroe C, Meekhof D~M, King B~E, Itano W~M and Wineland D~J 1995 {\em Phys. Rev. Lett.\/} {\bf 75} 4714--4717

\bibitem{bravyi2005universal}
Bravyi S and Kitaev A 2005 {\em Phys. Rev. A\/} {\bf 71} 022316

\bibitem{obrien2003demonstration}
O'Brien J~L, Pryde G~J, White A~G, Ralph T~C and Branning D 2003 {\em Nature\/} {\bf 426} 264--267

\bibitem{knill2001scheme}
Knill E, Laflamme R and Milburn G~J 2001 {\em Nature\/} {\bf 409} 46--52

\bibitem{kieling2010photonic}
Kieling K, O'Brien J~L and Eisert J 2010 {\em New J. Phys.\/} {\bf 12} 013003

\bibitem{franson2002highfidelity}
Franson J~D, Donegan M~M, Fitch M~J, Jacobs B~C and Pittman T~B 2002 {\em Phys. Rev. Lett.\/} {\bf 89} 137901

\bibitem{aharonovich2016solidstate}
Aharonovich I, Englund D and Toth M 2016 {\em Nat. Photon.\/} {\bf 10} 631--641

\bibitem{pelucchi2022potential}
Pelucchi E, Fagas G, Aharonovich I, Englund D, Figueroa E, Gong Q, Hannes H, Liu J, Lu C~Y, Matsuda N, Pan J~W, Schreck F, Sciarrino F, Silberhorn C, Wang J and J{\"o}ns K~D 2022 {\em Nat. Rev. Phys.\/} {\bf 4} 194--208

\bibitem{wang2020integrated}
Wang J, Sciarrino F, Laing A and Thompson M~G 2020 {\em Nat. Photon.\/} {\bf 14} 273--284

\bibitem{gea-banacloche2010impossibility}
{Gea-Banacloche} J 2010 {\em Phys. Rev. A\/} {\bf 81} 043823

\bibitem{imoto1985quantum}
Imoto N, Haus H~A and Yamamoto Y 1985 {\em Phys. Rev. A\/} {\bf 32} 2287--2292

\bibitem{matsuda2009observation}
Matsuda N, Shimizu R, Mitsumori Y, Kosaka H and Edamatsu K 2009 {\em Nat. Photon.\/} {\bf 3} 95--98

\bibitem{fushman2008controlled}
Fushman I, Englund D, Faraon A, Stoltz N, Petroff P and Vuckovic J 2008 {\em science\/} {\bf 320} 769--772

\bibitem{turchette1995measurement}
Turchette Q~A, Hood C~J, Lange W, Mabuchi H and Kimble H~J 1995 {\em Phys. Rev. Lett.\/} {\bf 75} 4710--4713

\bibitem{parigi2012observation}
Parigi V, Bimbard E, Stanojevic J, Hilliard A~J, Nogrette F, {Tualle-Brouri} R, Ourjoumtsev A and Grangier P 2012 {\em Phys. Rev. Lett.\/} {\bf 109} 233602

\bibitem{shi2022Highfidelityb}
Shi S, Xu B, Zhang K, Ye G~S, Xiang D~S, Liu Y, Wang J, Su D and Li L 2022 {\em Nat. Commun.\/} {\bf 13} 4454

\bibitem{thompson2017symmetryprotected}
Thompson J~D, Nicholson T~L, Liang Q~Y, Cantu S~H, Venkatramani A~V, Choi S, Fedorov I~A, Viscor D, Pohl T, Lukin M~D and Vuleti{\'c} V 2017 {\em Nature\/} {\bf 542} 206--209

\bibitem{khazali2019Polaritonb}
Khazali M, Murray C~R and Pohl T 2019 {\em Phys. Rev. Lett.\/} {\bf 123} 113605

\bibitem{browaeys2020many}
Browaeys A and Lahaye T 2020 {\em Nat. Phys.\/} {\bf 16} 132--142

\bibitem{ravets2014coherent}
Ravets S, Labuhn H, Barredo D, B{\'e}guin L, Lahaye T and Browaeys A 2014 {\em Nat. Phys.\/} {\bf 10} 914--917

\bibitem{sumarac2026controlling}
{\v S}umarac T, Qiu E~H, Tsesses S, Niu P, Menssen A~J, Xu W, Walther V, Deli{\'c} U, Choi S, Lukin M~D and Vuleti{\'c} V 2026 {\em arXiv:\,\rm 2601.06345 [atom-ph]\/}

\bibitem{shi2021quantum}
Shi X~F and Lu Y 2021 {\em Phys. Rev. A\/} {\bf 104} 012615

\bibitem{volz2006observation}
Volz J, Weber M, Schlenk D, Rosenfeld W, Vrana J, Saucke K, Kurtsiefer C and Weinfurter H 2006 {\em Phys. Rev. Lett.\/} {\bf 96} 030404

\bibitem{simon2007singlephoton}
Simon J, Tanji H, Ghosh S and Vuleti{\'c} V 2007 {\em Nat. Phys.\/} {\bf 3} 765--769

\bibitem{ritter2012elementary}
Ritter S, N{\"o}lleke C, Hahn C, Reiserer A, Neuzner A, Uphoff M, M{\"u}cke M, Figueroa E, Bochmann J and Rempe G 2012 {\em Nature\/} {\bf 484} 195--200

\bibitem{li2013Entanglementa}
Li L, Dudin Y~O and Kuzmich A 2013 {\em Nature\/} {\bf 498} 466--469

\bibitem{li2019semideterministic}
Li J, Zhou M~T, Yang C~W, Sun P~F, Liu J~L, Bao X~H and Pan J~W 2019 {\em Phys. Rev. Lett.\/} {\bf 123} 140504

\bibitem{sun2022deterministic}
Sun P~F, Yu Y, An Z~Y, Li J, Yang C~W, Bao X~H and Pan J~W 2022 {\em Phys. Rev. Lett.\/} {\bf 128} 060502

\bibitem{yang2020atomphoton}
Yang F, Liu Y~C and You L 2020 {\em Phys. Rev. Lett.\/} {\bf 125} 143601

\bibitem{ghosh2021creating}
Ghosh S, Rivera N, Eisenstein G and Kaminer I 2021 {\em Light Sci. Appl.\/} {\bf 10} 100

\bibitem{yang2022sequential}
Yang C~W, Yu Y, Li J, Jing B, Bao X~H and Pan J~W 2022 {\em Nat. Photon.\/} {\bf 16} 658--661

\bibitem{ye2023photonic}
Ye G~S, Xu B, Chang Y, Shi S, Shi T and Li L 2023 {\em Nat. Photon.\/} {\bf 17} 538--543

\bibitem{wilk2010EntanglementTwoIndividual}
Wilk T, Ga{\"e}tan A, Evellin C, Wolters J, Miroshnychenko Y, Grangier P and Browaeys A 2010 {\em Phys. Rev. Lett.\/} {\bf 104} 010502

\bibitem{isenhower2010demonstration}
Isenhower L, Urban E, Zhang X~L, Gill A~T, Henage T, Johnson T~A, Walker T~G and Saffman M 2010 {\em Phys. Rev. Lett.\/} {\bf 104} 010503

\bibitem{theis2016highfidelity}
Theis L~S, Motzoi F, Wilhelm F~K and Saffman M 2016 {\em Phys. Rev. A\/} {\bf 94} 032306

\bibitem{han2016implementing}
Han R, Ng H~K and Englert B~G 2016 {\em EPL (Europhysics Letters)\/} {\bf 113} 40001

\bibitem{muller2011prospects}
M{\"u}ller M~M, Haakh H~R, Calarco T, Koch C~P and Henkel C 2011 {\em Quantum Information Processing\/} {\bf 10} 771--792

\bibitem{goerz2014robustness}
Goerz M~H, Halperin E~J, Aytac J~M, Koch C~P and Whaley K~B 2014 {\em Phys. Rev. A\/} {\bf 90} 032329

\bibitem{saffman2020symmetrica}
Saffman M, Beterov I~I, Dalal A, P{\'a}ez E~J and Sanders B~C 2020 {\em Phys. Rev. A\/} {\bf 101} 062309

\bibitem{tian2015populationa}
Tian X~D, Liu Y~M, Cui C~L and Wu J~H 2015 {\em Phys. Rev. A\/} {\bf 92} 063411

\bibitem{su2017applications}
Su S~L, Tian Y, Shen H~Z, Zang H, Liang E and Zhang S 2017 {\em Phys. Rev. A\/} {\bf 96} 042335

\bibitem{li2018engineering}
Li D~X, Shao X~Q, Wu J~H, Yi X~X and Zheng T~Y 2018 {\em Opt. Express\/} {\bf 26} 2292

\bibitem{beterov2016twoqubit}
Beterov I~I, Saffman M, Yakshina E~A, Tretyakov D~B, Entin V~M, Bergamini S, Kuznetsova E~A and Ryabtsev I~I 2016 {\em Phys. Rev. A\/} {\bf 94} 062307

\bibitem{beterov2018adiabatic}
Beterov I~I, Hamzina G~N, Yakshina E~A, Tretyakov D~B, Entin V~M and Ryabtsev I~I 2018 {\em Phys. Rev. A\/} {\bf 97} 032701

\bibitem{idlas2016entanglement}
Idlas S, Domenzain L, Spreeuw R and Byrnes T 2016 {\em Phys. Rev. A\/} {\bf 93} 022319

\bibitem{zhao2017robusta}
Zhao Y~J, Liu B, Ji Y~Q, Tang S~Q and Shao X~Q 2017 {\em Sci. Rep.\/} {\bf 7} 16489

\bibitem{zhang2020submicrosecond}
Zhang C, Pokorny F, Li W, Higgins G, P{\"o}schl A, Lesanovsky I and Hennrich M 2020 {\em Nature\/} {\bf 580} 345--349

\bibitem{beterov2018fast}
Beterov I~I, Ashkarin I~N, Yakshina E~A, Tretyakov D~B, Entin V~M, Ryabtsev I~I, Cheinet P, Pillet P and Saffman M 2018 {\em Phys. Rev. A\/} {\bf 98} 042704

\bibitem{su2015simplified}
Su S~L, Guo Q, Wang H~F and Zhang S 2015 {\em Phys. Rev. A\/} {\bf 92} 022328

\bibitem{reiter2016scalable}
Reiter F, Reeb D and S{\o}rensen A~S 2016 {\em Phys. Rev. Lett.\/} {\bf 117} 040501

\bibitem{yang2021dissipative}
Yang C, Li D~X and Shao X~Q 2021 {\em Chinese Phys. B\/} {\bf 30} 023201

\bibitem{carr2013preparationa}
Carr A~W and Saffman M 2013 {\em Phys. Rev. Lett.\/} {\bf 111} 033607

\bibitem{li2020periodically}
Li R, Yu D, Su S~L and Qian J 2020 {\em Phys. Rev. A\/} {\bf 101} 042328

\bibitem{rao2014deterministic}
Rao D~D~B and M{\o}lmer K 2014 {\em Phys. Rev. A\/} {\bf 90} 062319

\bibitem{jau2016Entangling}
Jau Y~Y, Hankin A~M, Keating T, Deutsch I~H and Biedermann G~W 2016 {\em Nat. Phys.\/} {\bf 12} 71--74

\bibitem{young2021asymmetric}
Young J~T, Bienias P, Belyansky R, Kaufman A~M and Gorshkov A~V 2021 {\em Phys. Rev. Lett.\/} {\bf 127} 120501

\bibitem{yang2025entangling}
Yang C~W, Li J, Sun P~F, An Z~Y, Bao X~H and Pan J~W 2025 {\em Phys. Rev. Lett.\/} {\bf 135} 110802

\bibitem{PhysRevA.95.043429}
Shi X~F and Kennedy T~A~B 2017 {\em Phys. Rev. A\/} {\bf 95}(4) 043429

\bibitem{PhysRevA.97.033414}
Shi X~F and Kennedy T~A~B 2018 {\em Phys. Rev. A\/} {\bf 97}(3) 033414

\bibitem{ShiJPB2016}
Shi X~F, Svetlichnyy P and Kennedy T~A~B 2016 {\em J. Phys. B\/} {\bf 49} 074005

\bibitem{lampen2018Longliveda}
Lampen J, Nguyen H, Li L, Berman P~R and Kuzmich A 2018 {\em Phys. Rev. A\/} {\bf 98} 033411

\bibitem{zhang2011magicwavelength}
Zhang S, Robicheaux F and Saffman M 2011 {\em Phys. Rev. A\/} {\bf 84} 043408

\bibitem{wilson2022trapping}
Wilson J~T, Saskin S, Meng Y, Ma S, Dilip R, Burgers A~P and Thompson J~D 2022 {\em Phys. Rev. Lett.\/} {\bf 128} 033201

\bibitem{PhysRevLett.134.053604}
Jiao Y, Li C, Shi X~F, Fan J, Bai J, Jia S, Zhao J and Adams C~S 2025 {\em Phys. Rev. Lett.\/} {\bf 134}(5) 053604

\bibitem{shi2025coherence}
Shi X~F, Lu Y, Jiao Y and Zhao J 2025 {\em Phys. Rev. Appl.\/} {\bf 24} 044028

\bibitem{shi2020suppressing}
Shi X~F 2020 {\em Phys. Rev. Appl.\/} {\bf 13} 024008

\bibitem{bbv3-d4ch}
Shi X~F 2025 {\em Phys. Rev. A\/} {\bf 112}(4) 042401

\bibitem{sheng2017intracavity}
Sheng J, Chao Y, Kumar S, Fan H, Sedlacek J and Shaffer J~P 2017 {\em Phys. Rev. A\/} {\bf 96} 033813

\bibitem{guerlin2010cavity}
Guerlin C, Brion E, Esslinger T and M{\o}lmer K 2010 {\em Phys. Rev. A\/} {\bf 82} 053832

\bibitem{manetsch2025tweezer}
Manetsch H~J, Nomura G, Bataille E, Lv X, Leung K~H and Endres M 2025 {\em Nature\/} {\bf 647} 60--67

\bibitem{pichard2024rearrangement}
Pichard G, Lim D, Bloch {\'E}, Vaneecloo J, Bourachot L, Both G~J, M{\'e}riaux G, Dutartre S, Hostein R, Paris J, Ximenez B, Signoles A, Browaeys A, Lahaye T and Dreon D 2024 {\em Phys. Rev. Appl.\/} {\bf 22} 024073

\bibitem{radnaev2025universal}
Radnaev A, Chung W, Cole D and \emph{et al} 2025 {\em PRX Quantum\/} {\bf 6} 030334

\bibitem{kaufman2021quantum}
Kaufman A~M and Ni K~K 2021 {\em Nat. Phys.\/} {\bf 17} 1324--1333

\bibitem{han2018coherent}
Han J, Vogt T, Gross C, Jaksch D, Kiffner M and Li W 2018 {\em Phys. Rev. Lett.\/} {\bf 120} 093201

\bibitem{tu2022highefficiency}
Tu H~T, Liao K~Y, Zhang Z~X, Liu X~H, Zheng S~Y, Yang S~Z, Zhang X~D, Yan H and Zhu S~L 2022 {\em Nat. Photon.\/} {\bf 16} 291--296

\bibitem{borowka2024continuous}
Bor{\'o}wka S, Pylypenko U, Mazelanik M and Parniak M 2024 {\em Nat. Photon.\/} {\bf 18} 32--38

\bibitem{petrosyan2019microwave}
Petrosyan D, M{\o}lmer K, Fort{\'a}gh J and Saffman M 2019 {\em New J. Phys.\/} {\bf 21} 073033

\end{thebibliography}
\providecommand{\newblock}{}

\end{document}